\documentclass[12pt]{article}
\usepackage{amssymb,amsmath}
\usepackage{color}
\numberwithin{equation}{section}
\newtheorem{theorem}{Theorem}[section]     
\newtheorem{definition}[theorem]{Definition}

\newtheorem{proposition}[theorem]{Proposition}
\newtheorem{lemma}[theorem]{Lemma}
\newtheorem{de}[theorem]{Definition}

\newtheorem{corollary}[theorem]{Corollary}

\newtheorem{remark}[theorem]{Remark}

\def\d{\partial}

\def\p{\partial}
\def\n{\noindent}
\def\A{{\mathcal A}}
\def\f{\frac}

\def\proof{\noindent\hspace{2em}{\itshape Proof: }}
\def\QEDclosed{\mbox{\rule[0pt]{1.3ex}{1.3ex}}} 

\def\QED{\QEDclosed} 
\def\endproof{\hspace*{\fill}~\QED\par\endtrivlist\unskip}

\numberwithin{equation}{section}

\def\d{\partial}
\def\n{\noindent}
\def\f{\frac}
\def\proof{\noindent\hspace{2em}{\itshape Proof: }}
\def\QEDclosed{\mbox{\rule[0pt]{1.3ex}{1.3ex}}} 

\def\QED{\QEDclosed} 
\def\endproof{\hspace*{\fill}~\QED\par\endtrivlist\unskip}

\newcommand{\eqa}{\begin{eqnarray}}
\newcommand{\eeqa}{\end{eqnarray}}
\newcommand{\beq}{\begin{equation}}
\newcommand{\eeq}{\end{equation}}

\def\f{\frac}
\def\p{\partial}
\def\e{\epsilon}

\numberwithin{equation}{section}

\newtheorem{thm}{Theorem}

\begin{document}
\title{Reciprocal $F$-manifolds}
\author{Alessandro Arsie* and Paolo Lorenzoni**\\
{\small *Department of Mathematics and Statistics}\\
{\small University of Toledo,}
{\small 2801 W. Bancroft St., 43606 Toledo, OH, USA}\\
{\small **Dipartimento di Matematica e Applicazioni}\\
{\small Universit\`a di Milano-Bicocca,}
{\small Via Roberto Cozzi 53, I-20125 Milano, Italy}\\
{\small *alessandro.arsie@utoledo.edu,  **paolo.lorenzoni@unimib.it}}

\date{}
\maketitle
\vspace{-0.2in}

{\bf Abstract:} We consider the action of a special class of reciprocal transformation on the principal hierarchy
 associated to a semisimple $F$-manifold with compatible flat structure $(M,\circ,\nabla,e)$.
 Under some additional assumptions, the hierarchy obtained applying these reciprocal transformations is also associated to an $F$-manifold with compatible flat structure that we call reciprocal $F$-manifold. We also consider
 the special case of bi-flat $F$-manifolds $(M,\circ,*,\nabla^{(1)},\nabla^{(2)},e,E)$ and we study reciprocal
 transformations preserving flatness of both the connections $\nabla^{(1)}$ and  $\nabla^{(2)}$
  and how they act on corresponding solutions of an augmented Darboux-Egorov system.

\section{Introduction}
$F$-manifolds have been introduced by Hertling and Manin (see \cite{HM}) at the end of the nineties as a generalization of the framework of Frobenius manifolds built by Dubrovin. 
An $F$-manifold $(M, \circ)$ is a manifold equipped with a commutative, associative product $\circ$ on vector fields, satisfying a technical condition called Hertling-Manin condition. 
We will usually represent this product as a $(1,2)$-tensor field $c^i_{jk}$, so that the $i$-th component of the product vector field $X\circ Y$ is given by $(X\circ Y)^i=c^i_{jk}X^j Y^k$.
 
We first recall the following:
\begin{definition}[Manin, \cite{manin}]\label{flatFmanifold}
An $F$-manifold with compatible flat structure (flat $F$-manifold for short) is a manifold equipped with a flat torsionless connection $\nabla$ and a commutative associative product on vector fields $c^i_{jk}$ 
such that \beq\label{connectioninvariant}\nabla_l c^i_{jk}=\nabla_j c^i_{lk}. \eeq 
\end{definition}
In the previous definition there is no mention of the Hertling-Manin condition, since, in this case, it is automatically fulfilled. 

From now on we will focus our attention on {\em semisimple} flat $F$-manifolds, for which the product $\circ$ admits a flat identity, namely a vector field $e$ such that $e\circ X=X$ for each $X$ vector field and $\nabla e=0$. The adjective semisimple simply means that there exists a system of coordinates $(u^1, \dots u^n)$ (called {\em canonical coordinates} for the product) such that $c^i_{jk}=\delta^i_j \delta^i_k$ in these coordinates. Moreover,  the manifolds we consider are topologically open subsets of $\mathbb{R}^n$. 

Since in canonical coordinates $u^1, \dots, u^n$, the identity $e$ has necessarily the form $e=\sum_{i=1}^n \f{\p }{\p u^i}$, it is not difficult to see that the Christoffel symbols of $\nabla$ for a flat $F$-manifold $(M, \circ, \nabla, e)$ must satisfy the following conditions when expressed in canonical coordinates:
\begin{equation}
\label{nat-conn-eps}
\begin{aligned}
&\Gamma^i_{jk}=0\qquad\mbox{for $i\ne j\ne k\ne i$}\\
&\Gamma^i_{jj}=-\Gamma^i_{ji}\qquad\mbox{for $i\ne j$}\\
&\Gamma^i_{ii}=-\sum_{k\ne i}\Gamma^i_{ik}\ .
\end{aligned}
\end{equation}
Let us remark that the first and the second constraints in \eqref{nat-conn-eps} are equivalent to the compatibility of $\nabla$ with the product, while the second and the third requirements are equivalent to $\nabla e=0$. 
Obviously these constraints do not specify $\nabla$ uniquely; on the other hand, once the function $\Gamma^i_{ij}$ are given, the connection $\nabla$ is uniquely determined. We call a connection satisfying \eqref{nat-conn-eps} the \emph{natural connection} associated to the functions $\Gamma^i_{ij}$.

The relevance for the theory of integrable PDEs of the geometric framework provided by semisimple flat $F$-manifolds is twofold. On one hand, given a semisimple flat $F$-manifold $(M, \circ, \nabla, e)$ it has been proved in \cite{LPR} that a straightforward generalization of Dubrovin's principal hierarchy construction can be implemented.

On the other hand, as it has been proved in \cite{LP}, given a semi-Hamiltonian system (see below), it is possible to associate to it a semisimple $F$-manifold with compatible connection (in the general case, this connection will not be flat but it will satisfy an extended compatibility condition with respect to the product). For certain semi-Hamiltonian systems, like for instance the $\epsilon$-system, the associated semisimple $F$-manifold is indeed a flat $F$-manifold. 

Let us briefly recall how to associate a natural connection to a semi-Hamiltonian system. 
A hierarchy of quasilinear PDEs
\beq\label{qlPDEs}
u^i_t=v^i(u^1, \dots, u^n)u^i_x,\qquad i=1,\dots,n.
\eeq
is called {\em semi-Hamiltonian} if the characteristic velocities $v^i$, $i=1, \dots n$ satisfy the following system of PDEs:
\begin{equation}\label{SH}
\partial_j\left(\frac{\partial_k v^i}{v^i-v^k}\right)=
\partial_k\left(\frac{\partial_j v^i}{v^i-v^j}\right)\hspace{1
cm}\forall\, i\ne j\ne k\ne i.
\end{equation}
Once we have a semi-Hamiltonian system like \eqref{qlPDEs}, we define 
\beq\label{symmetries}
\Gamma^i_{ij}(u^1, \dots, u^n):=\f{\d_j v^i}{v^j-v^i}\qquad\mbox{for $i\ne j$}.
\eeq
A part from defining $\Gamma^i_{ij}$, equation \eqref{symmetries} can be thought of providing the equation for the characteristic velocities $w^1, \dots, w^n$ of the symmetries 
$$u^i_{\tau}=w^i(u^1, \dots, u^n)u^i_x, \quad i=1, \dots n,$$
of \eqref{qlPDEs} via
\beq\label{determinesymmetries}
\f{\p_j w^i}{w^i-w^j}=\f{\p_j v^i}{v^i-v^j}, \quad i\neq j.
\eeq
Observe that once $\Gamma^i_{ij}$ are given via the characteristic velocities using \eqref{symmetries}, the corresponding natural connection $\nabla$ \eqref{nat-conn-eps} is determined. Let us also remark that the semi-Hamiltonian condition, namely system \eqref{SH} provides the integrability conditions for the equations \eqref{determinesymmetries} that determine symmetries and also for the equation 
\beq\label{densityconservation}
\p_i \p_j A-\Gamma^i_{ij}\p_i A-\Gamma^j_{ji}\p_j A=0, \quad, i\neq j,
\eeq
whose solutions are exactly the conserved densities for \eqref{qlPDEs} (here $\Gamma^i_{ij}$ are given by \eqref{symmetries}). For more information about semi-Hamiltonian systems of hydrodynamic type see \cite{ts}. 

Let us point out that the natural connection $\nabla$ canonically associated to a semi-Hamiltonian system is not in general flat. However the following properties hold. 
The semi-Hamiltonian condition \eqref{SH} is equivalent to 
\begin{equation}\label{sh}
\d_i\Gamma^k_{kj}-\Gamma^k_{kj}\Gamma^j_{ij}
+\Gamma^k_{ik}\Gamma^k_{kj}-\Gamma^k_{ik}\Gamma^i_{ij}=0,\qquad\forall i\ne j\ne k\ne i
\end{equation}
which also gives the conditions
\begin{equation}
\label{shder}
\partial_j\Gamma^i_{ik}=\partial_k\Gamma^i_{ij},\qquad\forall i\ne j\ne k\ne i.
\end{equation}

In this paper we investigate a class of transformations among flat semisimple $F$-manifolds; these transformations arise from reciprocal transformations relating the corresponding semi-Hamiltonian PDEs. Originating from the study
 of gas dynamics, reciprocal transformations
 play an important role in the theory of integrable
 PDEs (see for instance \cite{RS,ts,F,P,FP,XZ,AG,A1})
 since they are non-trivial transformations preserving 
 the integrability. In particular they preserve the semi-Hamiltonian property and thus their action is well defined on natural connections. We will focus our attention on reciprocal transformations of the form:
\begin{equation}\label{RT}
d\tilde{t}=dt,\qquad d\tilde{x}=Adx+Bdt,
\end{equation}
preserving the time variable. The reason for restricting our attention to this class is due to the fact that they have a clear geometric interpretation at the level of flat $F$-manifolds. 
The functions $A$ and $B$ appearing above are the density and the current of a conservation law respectively. 

 The aim of the present work is to characterize  reciprocal transformations of the form \eqref{RT} preserving  
 flatness of the natural connection. We will show that such transformartions are generated by \emph{any} density of conservation law
 satisfying the complete compatible system:
\beq\label{flatnessreciprocal}
\begin{split}
\d_q\d_p A=\d_p A\,\Gamma^p_{pq}+\d_qA\,\Gamma^q_{pq}, \quad p\neq q\\
\d_p^2 A=-\sum_{l\ne p}\d_l\d_p A+\f{1}{A}\d_p\,A\sum_{l=1}^n\d_l\,A.
\end{split}
\eeq
Remarkably, {\em no extra conditions are necessary}. (To see that system \eqref{flatnessreciprocal} is complete, one defines $\theta_p:=\p_p \ln(A)$ and rewrite both equations in terms of $\theta_p$ and its first derivatives with respect to all the independent variables.)
Comparing this result with similar results obtained in the study of reciprocal transformations
 preserving the locality of Hamiltonian operators, we see that conditions \eqref{flatnessreciprocal} are much less stringent compared to those found in other works (see for instance  \cite{F}, \cite{A1}) and thus they provide a higher degree of flexibility. At least for reciprocal transformations of the special
 form \eqref{RT}, we believe that the framework provided by the point of view adopted here, emphasizing the role of natural connections rather than focusing on the flatness of the transformed metric,  seems more powerful.

The geometric counterpart of reciprocal transformations of the form \eqref{RT} sending a given flat natural connection to another flat natural connection is a map $\psi_A: (M,\circ,\nabla,e)\to(M,\circ,\tilde\nabla,e)$ between $F$-manifolds with compatible flat structure, which is explicitly given by  
\beq
\tilde\Gamma^i_{jk}=\Gamma^i_{jk}+c^i_{jk}e^l(\d_l\ln{A})-c^i_{lj}(\d_k\ln{A})e^l+c^l_{jk}(\d_l\ln{A})e^i-c^i_{lk}(\d_j\ln{A})e^l,
\eeq
where $\Gamma^i_{jk}$ and $\tilde \Gamma^i_{jk}$ are the Christoffel symbols of $\nabla$ and $\tilde \nabla$ respectively. 
In this geometric set-up the system \eqref{flatnessreciprocal} for $A$ can be rewritten as
\beq\label{geometricflatnessreciprocal}
\nabla_q\nabla_p A=e(\ln{A})c^l_{pq}\nabla_l A,
\eeq
where $p$ and $q$ are possibly equal and the following definition is completely natural:
\begin{definition}
We call two flat $F$-manifolds $(M,\circ,\nabla,e)$, $(M,\circ,\tilde\nabla,e)$ related via a $\psi_A$ for some $A$ satisfying
 \eqref{geometricflatnessreciprocal}
reciprocal $F$-manifolds. 
\end{definition}

The paper is organized as follows. In Section 2 we recall few facts about reciprocal transformations of the form \eqref{RT}, we derive conditions \eqref{flatnessreciprocal} and prove that the corresponding system is complete and compatible. In Section 3 we provide a geometric interpretation within the framework of flat $F$-manifolds. We also consider transformations of the form \eqref{RT} sending a bi-flat $F$-manifold to another bi-flat $F$-manifold (bi-flat $F$-manifolds have been introduced by the authors in \cite{AL2}, and they have remarkable properties in terms of recurrence relations, see \cite{AL1}). It turns out that such transformations are generated by homogeneous
 flat coordinates depending only on the differences of the canonical coordinates.
 In Section 4 we determine the orbit structure of all semi-Hamiltonian systems that can be obtained from a given diagonal semi-Hamiltonian system whose associated natural connection is flat, iterating reciprocal transformation of the form \eqref{RT} preserving the flat $F$-manifold structure. 
 In Section 5, we further explore the action of reciprocal transformations of the form \eqref{RT} on bi-flat $F$-manifolds.
 We briefly recall from \cite{AL2} how it is possible to associate to a solution of a suitably augmented Darboux-Egorov system,
 a bi-flat $F$-manifold structure and we explore how the reciprocal transformations considered in this paper act on solutions
 of this augmented Darboux-Egorov system. In particular, we prove that if one starts with a bi-flat $F$-manifold constructed
 from a solution of this Darboux-Egorov system, also the new bi-flat $F$-manifold one obtains is associated to a (different) solution
 of the extended Darboux-Egorov system.
In Section 6 we review few facts about a particular semi-Hamiltonian system, the so called $\epsilon$-system, whose natural connection is flat.
 Actually this system is associated to a bi-flat $F$-manifold structure. We will use it in Sections 7, 8 to construct explicit examples of reciprocal transformations preserving the flatness of the natural connection in dimension 2 and 3 respectively. Moreover, we will use it in Section 9 to build explicit examples of reciprocal transformations preserving the bi-flat $F$-manifold structure. 

\section{Reciprocal transformations preserving flatness}

Let us consider reciprocal transformations, namely a change of independent coordinates of the form
\beq\label{RT3}
d\tilde{t}=dt,\qquad d\tilde{x}=Adx+Bdt.
\eeq
By definition, since $d\tilde{x}$ is an exact form, we have $d\left(Adx+Bdt\right)=0$. This means 
 that $A$ is a density of conservation law and $B$
 the related current. Indeed, from $d\left(Adx+Bdt\right)=0$ we have $A_t dt\wedge dx+B_x dx\wedge dt=0$, or $A_t=B_x$. 
 In other words $$A_t=\sum_l\d_l A u^l_t=\sum_{l=1}^n\d_l A v^l u^l_x$$
has to be a total $x$-derivative. It is well-known that this is equivalent to
\beq
\f{\delta}{\delta u^i}\left(\sum_l\d_l A v^l u^l_x\right)=0,
\eeq
where $\f{\delta}{\delta u^i}$ is the variational derivative. 
 After having performed some straightforward computations, the previous condition on $A$ is equivalent to 
$$\d_i\d_j A
-\d_j A\,\Gamma^j_{ij}-\d_i A\,\Gamma^i_{ij}=0,\quad i\neq j$$
which is exactly the condition for $A$ being a density of conservation law. 
It is easy to check that as effect of this transformation the characteristic velocities transform as:
$$v^i\to\tilde{v}^i(u)=Av^i-B,\,\,\,i=1,\dots,n.$$
Indeed, using \eqref{RT3} we find
$$u^i_t=u^i_{\tilde t}\f{d \tilde t}{d t}+u^i_{\tilde x}\f{d\tilde x}{d x}=u^i_{\tilde t}+Bu^i_{\tilde x},$$
$$u^i_x=u^i_{\tilde x}\f{d\tilde x}{dx}=A u^i_{\tilde x},$$
and since $u^i_t=v^i u^i_x$, and $u^i_{\tilde t}=\tilde v^i u^i_{\tilde x}$, we obtain immediately the transformation of the characteristic velocities. \begin{lemma}\label{transChristoffel}
As a consequence of $v^i\to\tilde{v}^i(u)=Av^i-B,\,\,\,i=1,\dots,n,$ the Christoffel symbols $\Gamma^i_{ij}:=\f{\d_j v^i}{v^j-v^i}$ transform as
\beq\label{transChristoffelformula}\Gamma^i_{ij}\to\tilde\Gamma^i_{ij}=\Gamma^i_{ij}
-\d_j\,\ln{A}.\eeq
\end{lemma}
\proof
Since $\tilde \Gamma^i_{ij}:=\f{\p_j \tilde v^i}{\tilde v^j-\tilde v^i}$, by direct substitution we have
$$
\tilde \Gamma^i_{ij}=\Gamma^i_{ij}+\f{v^i \p_j A}{A(v^j-v^i)}-\f{\p_j B}{A(v^j-v^i)}.
$$
The fact that $A_t=B_x$ implies that the following system of PDEs hold identitcally: $\p_j B-v^j\p_j A=0$ (no sum over $j$). 
Therefore substituting $\p_jB=v^j\p_j A$ in the previous expression we get immediately \eqref{transChristoffelformula}.
\endproof
The new characteristic velocities still satisy the semi-Hamiltonian condition, so the resulting system $u^i_{\tilde t}=(Av^i-B)u^i_{\tilde x}$ of hydrodynamic type is also semi-Hamiltonian. 

Suppose now that the natural connection $\nabla$ associated to functions $\Gamma^i_{ij}$ is flat. Let us consider the natural connection $\tilde{\nabla}$ associated to functions
 $\tilde\Gamma^i_{ij}$:
\begin{equation}
\label{nat-conn-eps2}
\begin{aligned}
&\tilde\Gamma^i_{jk}=0\qquad\mbox{for $i\ne j\ne k\ne i$}\\
&\tilde\Gamma^i_{jj}=-\tilde\Gamma^i_{ji}=\Gamma^i_{jj}+\d_j\ln{A}\qquad\mbox{for $i\ne j$}\\
&\tilde\Gamma^i_{ji}=\Gamma^i_{ji}-\d_j\ln{A}\qquad\mbox{for $i\ne j$}\\
&\tilde\Gamma^i_{ii}=-\sum_{k\ne i}\tilde\Gamma^i_{ik}=\Gamma^i_{ii}+\sum_{l\ne i}\d_l\ln{A}\ .
\end{aligned}
\end{equation}
Now we determine under which conditions on $A$ the deformed natural connection $\tilde{\nabla}$ is flat, assuming that the initial natural connection $\nabla$ is flat.
\begin{thm}
The transformed natural connection $\tilde{\nabla}$ is flat iff the density $A$ of conservation law satisfies the equation:
\beq\label{transformedflat2}
\sum_{l=1}^n\d_k\d_l\,A=\f{1}{A}\d_k\,A\sum_{l=1}^n\d_l\,A\eeq
\end{thm} 

\n
\proof In order to get a new flat connection $\tilde{\nabla}$ we have to impose
\begin{eqnarray*}
\tilde R^i_{iki}&=&\d_k\tilde\Gamma^i_{ii}
-\d_i\tilde\Gamma^i_{ik}\\
\tilde R^i_{qqi}&=&\d_q\tilde\Gamma^i_{qi}
-\d_i\tilde\Gamma^i_{qq}
+\tilde\Gamma^i_{iq}(\tilde\Gamma^i_{iq}-\tilde\Gamma^q_{iq})
-\sum_{p\ne i,q}\tilde\Gamma^i_{pi}\tilde\Gamma^p_{qq}
-\tilde\Gamma^i_{ii}\tilde\Gamma^i_{qq}
-\tilde\Gamma^i_{qi}\tilde\Gamma^q_{qq},
\end{eqnarray*}
since, for a natural connection, these are the only possibly non-vanishing components of the Riemann tensor. 
The first condition gives us
\begin{eqnarray*}
&&0=\d_k\tilde\Gamma^i_{ii}-\d_i\tilde\Gamma^i_{ik}=
\d_k\sum_{l\ne i}\d_l\ln{A}+\d_i\d_k\ln{A}=\\
&&\d_k\left(\sum_{l=1}^n\d_l\ln{A}\right)=\d_k\left(\f{\sum_{l=1}^n\d_l\,A}{A}\right)=
\f{\sum_{l=1}^n\d_k\d_l\,A}{A}-\f{\d_k\,A\sum_{l=1}^n\d_l\,A}{A^2}
\end{eqnarray*}
that is, multiplying by $A$:
$$\sum_{l=1}^n\d_k\d_l\,A=\f{\d_k\,A\sum_{l=1}^n \d_l\,A}{A},$$
which can be written as 
$$\p_k e(A)=\f{e(A)\p_k A}{A},$$
or equivalently 
\beq\label{aiuto9999}\p_k (e(\ln(A))=0.\eeq

The second condition is automatically satisfied since $A$ is a density of conservation law. Indeed we have 
\begin{eqnarray*}
&&\tilde R^i_{qqi}= -\d^2_q\ln{A}-\d_i\d_q\ln{A}
-\d_q\ln{A}(\Gamma^i_{iq}-\Gamma^q_{iq})
-\Gamma^i_{iq}(\d_q\ln{A}-\d_i\ln{A})+\\
&&+\d_q\ln{A}(\d_q\ln{A}-\d_i\ln{A})
-\sum_{l\ne i,q}\d_l\ln{A}\,\Gamma^l_{lq}
-\sum_{l\ne i,q}\Gamma^i_{li}\d_q\ln{A}+\\
&&+\sum_{l\ne i,q}\d_l\ln{A}\,\d_q\ln{A}
+\sum_{l\ne i}\d_l\ln{A}\,\Gamma^i_{iq}
+\sum_{l\ne i}\Gamma^i_{il}\d_q\ln{A}
-\sum_{l\ne i}\d_l\ln{A}\,\d_q\ln{A}+\\
&&-\sum_{l\ne q}\d_q\ln{A}\,\Gamma^q_{lq}
-\sum_{l\ne q}\Gamma^i_{qi}\d_l\ln{A}
+\sum_{l\ne q}\d_q\ln{A}\,\d_l\ln{A},\\
\end{eqnarray*}
where we used the fact that the undeformed connection $\nabla$ is indeed flat. 
Then using also the condition on $A$ coming from $\d_k\tilde\Gamma^i_{ii}-\d_i\tilde\Gamma^i_{ik}=0$ written above, we obtain that the previous expression for $\tilde R^i_{qqi}$ is equal to
\begin{eqnarray*}
&&\d_q\left(\sum_{l\ne i,q}\d_l\ln{A}\right)
+\d_q\ln{A}\,\Gamma^q_{iq}-\d_q\ln{A}\,\d_i\ln{A}+\\
&&-\sum_{l\ne i,q}\d_l\ln{A}\,\Gamma^l_{lq}
-\sum_{l\ne q}\d_q\ln{A}\,\Gamma^q_{lq}
+\sum_{l\ne q}\d_q\ln{A}\,\d_l\ln{A}=\\
&&=\f{1}{A}\sum_{l\ne i,q}\left(\d_q\d_l A
-\d_l A\,\Gamma^l_{lq}
-\d_q A\,\Gamma^q_{lq}\right)=0,
\end{eqnarray*}
where the last equality comes from \eqref{densityconservation}.
\endproof

\begin{corollary}\label{simplecorollary}
The transformed natural connection $\tilde{\nabla}$ is flat iff $A$ is a density of conservation law and $A$ satisfies 
\beq\label{transformedflat3}
e(A)=hA,
\eeq
where $e=\sum_{l=1}^n \f{\p }{\p u^l}$ and $h$ is an arbitrary constant. 
\end{corollary}
\proof Immediate because of \eqref{aiuto9999}. \endproof

Combining equation \eqref{transformedflat2} with equation \eqref{densityconservation}, we obtain a compatible system involving all second derivatives of $A$. This is the content of the following:
\begin{thm}
The system 
\begin{equation}
\label{Asystem}
\begin{aligned}
\d_q\d_p A&=&\d_p A\,\Gamma^p_{pq}+\d_qA\,\Gamma^q_{pq}, \quad p\neq q\\
\d_p^2 A&=&-\sum_{l\ne p}\d_l\d_p A+\f{\d_p\,A\sum_{l}\d_l\,A}{A}
\end{aligned}
\end{equation}
is compatible.
\end{thm}

\n
\emph{Proof}. The compatibility conditions
$$\d_p\d_q\d_r A-\d_q\d_p\d_r A =0$$
for distinct indices follow from the condition \eqref{sh}. We have only to check
$$\d_p\d_q\d_p A-\d_q\d_p^2 A=0.$$
The following identities holds because of system \eqref{Asystem}
\begin{eqnarray*}
\d_p\d_q\d_p A&=&\d^2_p A\,\Gamma^p_{pq}+\d_p A\,\d_p\Gamma^p_{pq}
+\d_p\d_qA\,\Gamma^q_{pq}+\d_qA\,\d_p\Gamma^q_{pq},\\
\d_q\d_p^2 A&=&\textcolor{red}{-}\sum_{l\ne p}\d_l\d_q\d_p A+\p_q\left(\d_p\,A\sum_{l}\d_l\,\ln(A) \right)\\
&=&-\sum_{l\ne p}\d_l\left(\d_p A\,\Gamma^p_{pq}+\d_qA\,\Gamma^q_{pq}\right)
+\d_q\d_p\,A\sum_{l}\d_l\,\ln{A},
\end{eqnarray*}
where in the last equality in the second equation we used the fact that $e(\ln(A))$ is a constant. 
Combining the two previous expressions we obtain
\beq\label{aiuto3333}
\begin{split}
\d_p\d_q\d_p A-\d_q\d_p^2 A=\sum_l\d_l\d_p A\,\Gamma^p_{pq}+\d_p A\sum_l\d_l\Gamma^p_{pq}
+\sum_l\d_l\d_qA\,\Gamma^q_{pq}+\\
+\d_qA\sum_l\d_l\Gamma^q_{pq}
-\d_q\d_p\,A\sum_{l}\d_l\,\ln{A}=0.
\end{split}
\eeq
Indeed
\begin{eqnarray*}
&&\sum_l \d_l\Gamma^i_{ik}=\sum_{l\ne i,k} \d_l\Gamma^i_{ik}+\d_i\Gamma^i_{ik}+\d_k\Gamma^i_{ik}=\sum_{l\ne i,k} \d_k\Gamma^i_{il}+\d_i\Gamma^i_{ik}+\d_k\Gamma^i_{ik}
\end{eqnarray*}
using \eqref{shder} in the last equality. Furthermore we have 
\begin{eqnarray*}
&&\sum_{l\ne i,k} \d_k\Gamma^i_{il}+\d_i\Gamma^i_{ik}+\d_k\Gamma^i_{ik}=\\
&&-\d_k\Gamma^i_{ik}-\d_k\Gamma^i_{ii}
+\d_i\Gamma^i_{ik}
+\d_k\Gamma^i_{ik}=0,
\end{eqnarray*}
where the first equality comes from $\Gamma^i_{ii}=\sum_{l\neq i}\Gamma^i_{il}$ (the third of \eqref{nat-conn-eps})
 while the second equality is due to \eqref{shder}. 
Moreover we have 
\begin{eqnarray*}
&&\sum_l\d_l\d_p A\,\Gamma^p_{pq}
+\sum_l\d_l\d_qA\,\Gamma^q_{pq}-\d_q\d_p\,A\sum_{l}\d_l\,\ln{A}=\\
&&\sum_{l}\d_l\,\ln{A}\left(\d_p\,A\,\Gamma^p_{pq}+\d_q\,A\,\Gamma^q_{pq}-\d_q\d_p\,A\right)=0.
\end{eqnarray*}
Therefore \eqref{aiuto3333} is indeed satisfied and compatibility holds.
\endproof
\begin{remark}
System \eqref{Asystem} can be seen to complete if one rewrites it with respect to $\theta_p:=\p_p \ln(A)$. Indeed, written with respect to $\theta_p$, $p=1, \dots, n$ it is equal to 
\beq
\begin{split}
\p_q \theta_p=\theta_p\Gamma^p_{pq}+\theta_q \Gamma^q_{qp}-\theta_p \theta_q, \quad p\neq q,\\
\p_p \theta_p=-\theta_p^2+\theta_p\sum_{l=1}^n \theta_l-\sum_{l\neq p}\left(\p_l \theta_p+\theta_p\theta_l \right).
\end{split}
\eeq
\end{remark}
Before concluding this section, let us observe that once a density of conservation law satisfying \eqref{transformedflat3} is computed, to determine the corresponding conserved current $B$ one has to solve the linear system of PDEs
$$\p_i B-v^i \p_i A=0, \quad i=1, \dots, n,$$
for the unknown $B$. This system for $B$ is automatically compatible if $A$ is a density of conservation law and if $\Gamma^i_{ij}=\f{\d_j v^i}{v^j-v^i}$.

\section{Geometric interpretation}
In this section, we give a geometric interpretation of system \eqref{Asystem} within the framework of flat $F$-manifolds. We also consider reciprocal transformations acting on two flat connections, this time within the set-up of bi-flat $F$-manifolds. 

As pointed out in the Introduction, to any semi-Hamiltonian system one can associate a natural connection, which, albeit not always flat, satisfies a suitable extended compatibility condition with respect to the product. In \cite{LPR} a distinguished class of $F$-manifolds was identified, corresponding exactly to semi-Hamiltonian systems.  
This class is given by the following definition:
\begin{de}
\label{defi:fmancc}
An \emph{$F$-manifold with compatible connection} is a manifold endowed with an associative commutative multiplicative structure
 $\circ$ on vector fields given by a $(1,2)$-tensor field $c$ and a torsionless connection $\nabla$ satisfying condition 
\beq\label{sccinv}
\left(\nabla_X \circ\right)\left(Y,Z\right)=\left(\nabla_Y \circ\right)\left(X,Z\right),
\eeq
 and condition
\beq\label{rc-intri}
Z\circ R(W,Y)(X)+
W\circ R(Y,Z)(X)+Y\circ R(Z,W)(X)=0,
\eeq
where $R(W,Y)$ is the curvature tensor of $\nabla$, 
for any choice of the vector fields $(X,Y,W,Z)$.
In local coordinates, condition \eqref{sccinv} is just \eqref{connectioninvariant} while the other condition reads 
\beq\label{shc}
R^k_{lmi}c^n_{pk}+R^k_{lip}c^n_{mk}+R^k_{lpm}c^n_{ik}=0\ .
\eeq
\end{de}
Notice that if $\nabla$ is flat then condition \eqref{rc-intri} is automatically fulfilled, so flat $F$-manifolds are a special subclass of $F$-manifolds with compatible connection. 

In the semisimple case condition \eqref{shc} reduces to semi-Hamiltonian condition \eqref{SH}. This means that reciprocal
 transformations act naturally on $F$-manifold with compatible connection. However only part of the Christoffel symbols 
 are prescribed by the transformation and even the transformation law for the product is unclear. This is due to the fact that reciprocal
 transformations are well defined on semi-Hamiltonian hierarchies and the same semi-Hamiltonian hierarchy can be associated to {\em different}
 $F$-manifolds. 
 
 To avoid this ambiguity one has 
 to introduce equivalence classes of semisimple $F$-manifolds with compatible connection. We give the following: 
 \begin{definition}\label{definitionFequivalent} Let $(M, \circ, \nabla, e)$ and $(M, \star, \tilde\nabla, E)$ be two semisimple $F$-manifolds with compatible connections $\nabla$ and $\tilde \nabla$ respectively, which we assume equipped with identities $e$ and $E$. (We do not assume necessarily $\nabla e=0$ and $\tilde \nabla E=0$). Assume also that $E$ is an eventual identity (see \cite{manin}) relating $\circ$ and $\star$, which means that $E$ is invertible\footnote{There is also an additional requirement
 which is equivalent, in the semisimple case, to the vaishing of the Nijenhuis torsion of $E\circ$.} with respect to $\circ$ and $X\star Y:=X\circ Y \circ E^{-1}$. Then we say that $(M, \circ, \nabla, e)$ and $(M, \star, \tilde\nabla, E)$ are {\em equivalent} if $\nabla$ and $\tilde\nabla$ are almost hydrodynamically equivalent (see \cite{AL1}), namely 
 if $\Gamma^i_{ij}=\tilde \Gamma^i_{ij}$, $i\neq j$, in the canonical coordinates for $\circ$ or in the canonical coordinates for $\star$. 
 \end{definition}
 Since almost hydrodynamically equivalent connections define the same semi-Hamiltonian system, the Definition above identifies an equivalence class of $F$-manifold with compatible connection having the same associated hierarchy. 
 
 The Christoffel symbols $\Gamma^i_{ij}$ fix uniquely an equivalence class and therefore the reciprocal transformation is well defined on the equivalence class. In practice, one can fix a representative in each equivalence class and define the transformation on these representatives. For instance, one can fix the product $\circ$
 and choose the connection $\nabla$ compatible with this product and satisfying the additional requirement $\nabla e=0$ (where $e$ is the unity vector field). This
 is just the natural connection we introduced at the beginning of the paper.

In this work we restrict our investigation to the special case where the natural connection is flat, so we are interested in $F$-manifold with flat compatible connection, for brevity flat $F$-manifold. The notion of flat $F$-manifold was recalled in Definition \ref{flatFmanifold} in the Introduction. 
Using commutativity of the algebra, it is easy to check  that in flat coordinates we have
$$c^i_{jk}=\d_j\d_k C^i.$$
Let us remark that in the Frobenius manifold, due to the existence of metric $g$ compatible with $\nabla$ and invariant with respect to $\circ$, one can make an additional step and obtain $C^i=\eta^{ij}\d_j F$ for a suitable function $F$ (the Frobenius potential).

The following Theorem provides a geometric interpretation of a reciprocal transformation of the form \eqref{RT} using flat $F$-manifolds. First we need a simple 
\begin{lemma}\label{aiuto333}
Equations \eqref{flatnessreciprocal} expressing the fact that $A$ is a density of conservation law preserving flatness of the natural connection are equivalent to 
\beq\label{flatnessreciprocal4}\nabla_q\nabla_pA=e(\ln{A})c^l_{pq}\nabla_l A\eeq
\end{lemma}
\proof
For $p\neq q$ the right hand side of \eqref{flatnessreciprocal4} is identically zero since $c^l_{pq}=\delta^l_p\delta^l_q$, therefore one has $\nabla_q \nabla_p A=0$ which is exactly the first of \eqref{flatnessreciprocal}, saying that $A$ is a density of conservation law. For $p=q$, \eqref{flatnessreciprocal4} reduces to 
$$\nabla_p (\p_p A)=\f{1}{A}e(A)\p_p A.$$
Expanding $\nabla_p (\p_p A)$ one obtains 
$$\nabla_p (\p_p A)=\p^2_p A-\Gamma^p_{pp}\p_p A-\sum_{l, l\neq p}\Gamma^l_{pp}\p_l A.$$
Using the properties of the natural connection \eqref{nat-conn-eps}, the previous expression becomes 
$$\nabla_p (\p_p A)=\p^2_p A+\sum_{l, l\neq p}\Gamma^p_{pl}\p_p A+\sum_{l, l\neq p}\Gamma^l_{lp}\p_l A.$$
Finally using the equation for $A$ being a density of conservation law, we have that $\sum_{l, l\neq p}\Gamma^p_{pl}\p_p A+\sum_{l, l\neq p}\Gamma^l_{lp}\p_l A$ is equal to $\sum_{l, l\neq p} \p_l \p_p A$, so that 
$$\nabla_p (\p_p A)=\p^2_p A +\sum_{l, l\neq p} \p_l \p_p A.$$
Substituting this into $\nabla_p (\p_p A)=\f{1}{A}e(A)\p_p A$ and rearranging terms one recognize the second equation in \eqref{flatnessreciprocal}. 
\endproof
\begin{thm}\label{conditionsflatintrinsic}
Let $(M,\nabla,\circ,e)$ a (semisimple) $F$-manifold with compatible flat structure and flat identity $e$.  Given a function $A$ satisfying
\beq\label{sysomega}
\nabla_q\nabla_p A=e(\ln{A})c^l_{pq}\nabla_l A,
\eeq
(with $p$ possibly equal to $q$)
the quadruple  $(M,\tilde\nabla,\circ,e)$ with connection $\tilde\nabla$  whose Christoffel symbols are given by the formula:
\beq\label{tildenabla}
\tilde\Gamma^i_{jk}=\Gamma^i_{jk}+c^i_{jk}e^l\omega_l-c^i_{lj}\omega_ke^l+c^l_{jk}\omega_le^i-c^i_{lk}\omega_je^l, \quad \omega_l:=\nabla_l(\ln( A))
\eeq
define a new  (semisimple) $F$-manifold with compatible flat structure and flat identity $e$.
\end{thm}
\proof One can easily check, by straightforward 
 computations, that conditon \eqref{tildenabla} reduces
 to \eqref{nat-conn-eps2}.  The fact that \eqref{sysomega} reduces to \eqref{Asystem} is just the content of the Lemma \ref{aiuto333}. 
\endproof

Let us observe that the two reciprocal flat $F$-manifold $(M,\nabla,\circ,e)$ and $(M,\tilde\nabla,\circ,e)$ constructed via Theorem \ref{conditionsflatintrinsic} are not equivalent according to Definition \ref{definitionFequivalent} (it is not forbidden to consider $E:=e$ in that Definition), since they are associated to different semi-Hamiltonian systems.

\begin{remark}
It is quite natural to ask what happens for more general reciprocal transformations involving also changes in
 the time variable:
\beq
d\tilde{x}=Adx+Bdt,\qquad d\tilde{t}=Mdx+Ndt.
\eeq
Unfortunately in this case the transformation law for the Christoffel symbols is much more complicated and it is given by 
\begin{eqnarray*}
\tilde{\Gamma}^i_{ij}=\left[\f{N-Mv^j}{N-Mv^i}\right]\Gamma^i_{ij}+\f{\d_j A(N-Mv^j)}{BM-AN}+\f{\d_j M(Av^i-B)}{BM-AN}\left[\f{N-Mv^j}{N-Mv^i}\right]. 
\end{eqnarray*}
Even if one considers the transformation of time alone, thus setting 
$A=1,\,B=0$, it is not clear how to give an interpretation of the modified connection in terms of the geometry of a flat $F$-manifold. 
\end{remark} 

In the last part of this Section, we focus our attention on reciprocal transformations of the form \eqref{RT} preserving a bi-flat $F$-manifold structure. Bi-flat $F$-manifolds have been introduced in \cite{AL2} as a convenient generalization of the concept of (conformal) Frobenius manifolds, particularly adapted to study integrable PDEs of hydrodynamic type. 
Let us recall the relevant definition. 
\begin{definition}\label{defibiflat}
A \emph{bi-flat} semisimple $F$-manifold $(M,\nabla^{(1)},\nabla^{(2)},\circ,*,e,E)$
 is a semisimple $F$-manifold $(M,\circ,e)$ endowed with a pair
 of flat connections $\nabla^{(1)}$ and $\nabla^{(2)}$ and with an eventual identity $E$ satisfying the following conditions:
\begin{itemize}
\item $\nabla^{(1)}$ is compatible with the product $\circ$ 
and $\nabla^{(1)} e=0$,
\item $\nabla^{(2)}$ is compatible with the product $*$ and $\nabla^{(2)} E=0$,
\item $\nabla^{(1)}$ and $\nabla^{(2)}$ are almost hydrodynamically equivalent, namely in canonical coordinates for $\circ$ one has
 $\Gamma^{(1)i}_{ij}-\Gamma^{(2)i}_{ij}=0$, $i\neq j$, where $\Gamma^{(1)}$ and $\Gamma^{(2)}$
 are the Christoffel symbols associated to the connections $\nabla^{(1)}$ and $\nabla^{(2)}$ respectively. An analogous statement holds
 in canonical coordinates for $*$,  see \cite{AL1}.
\end{itemize}
\end{definition}

Observe that any Frobenius manifold possesses automatically a bi-flat $F$-manifold structure in the above sense. The second flat connection $\nabla^{(2)}$
 is hydrodymically equivalent \footnote{The difference between the Christoffel symbols
 is proportional to the structure constants.} to
 the Levi-Civita connection $\nabla$ of the intersection form
 which, as it is well-known, does not fulfill, in general,  the condition $\nabla E=0$.
 
 Let us recall from \cite{AL2} that the data $E, *, \nabla^{(2)}$ have the following expressions in the canonical coordinates of the first product $\circ$:
 $$E:=\sum_{i=1}^n u^i \f{\p }{\p u^i}, \quad c^{*i}_{jk}=\f{1}{u^i}\delta^i_j \delta^i_k,$$
 where $c^{*i}_{jk}$ are the structure constants of $\star$ and $\nabla^{(2)}$ has Christoffel symbols:
 \beq\label{dualconnection}
 \begin{split}
 \Gamma^{(2)i}_{jk}&:=0\qquad\forall i\ne j\ne k \ne i,\\
\Gamma^{(2)i}_{ij}&:=\Gamma^{(1)i}_{ij}\\
\Gamma^{(2)i}_{jj}&:=-\f{u^i}{u^j}\Gamma^{(2)i}_{ij}\qquad i\ne j,\\
\Gamma^{(2)i}_{ii}&:=-\sum_{l\ne i}\f{u^l}{u^i}\Gamma^{(2)i}_{li}-\f{1}{u^i}.
 \end{split}
 \eeq
To determine how $\Gamma^{(2)i}_{jk}$ transform under the action of the reciprocal transformation generated by $A$ we proceed as in \eqref{nat-conn-eps2} and we find the following formulas, valid in the canonical coordinates of $\circ$:
 \beq\label{transformationdual}
\begin{split}
 \tilde \Gamma^{(2)i}_{jk}=0,
 \quad i\neq j\neq k\neq i,\\
 \tilde \Gamma^{(2)i}_{jj}=\Gamma^{(2)i}_{ jj}+\f{u^i}{u^j}\p_j \ln(A), \quad i\neq j,\\
 \tilde \Gamma^{(2)i}_{ii}=\Gamma^{(2)i}_{ii}+\sum_{l\, l\neq i}\f{u^l}{u^i}\p_l \ln(A).
 \end{split}
 \eeq
 
 Summarizing, we have the following Proposition which is the counterpart of Theorem \ref{conditionsflatintrinsic}.
 \begin{proposition}
 The conditions for a function $A$ to map via $\psi_A$ a flat semisimple $F$-manifold $(M, \nabla^{(2)}, *, E)$ to another flat semisimple $F$-manifold $(M, \tilde{\nabla}^{(2)}, *, E)$ are given by 
 \beq\label{secondconnectionpreserved}
 \nabla^{(2)}_{q}\nabla^{(2)}_{p} A=E(\ln{A})c^{*l}_{pq}\nabla^{(2)}_{l} A.
 \eeq
 Moreover the  connection $\nabla^{(2)}$ transforms as
\beq\label{transformationdualin}
\tilde\Gamma^{(2)i}_{jk}=\Gamma^{(2)i}_{jk}+c^{*i}_{jk}E^l\omega_l-c^{*i}_{lj}\omega_kE^l+c^{*l}_{jk}\omega_lE^i-c^{*i}_{lk}\omega_jE^l, \quad \omega_l:=\nabla^{(2)}_l(\ln( A)).
\eeq
\end{proposition}
 \proof
 Since in canonical coordinates for $*$, $(w^1, \dots, w^n)$ we have $E=\sum_{i=1}^n \f{\d}{\d w^i}$, $c^{*i}_{jk}=\delta^i_j \delta^i_k$,  $\nabla^{(2)}$ has just the form of the natural \eqref{nat-conn-eps} connection in such coordinates. Then condition \eqref{secondconnectionpreserved} follows immediately from Theorem \ref{conditionsflatintrinsic}. Also since in the canonical coordinates of $\circ$, $E^l=u^l$ and $c^{*i}_{lj}=\f{1}{u^i}\delta^i_l \delta^i_j$, we see that the formula \eqref{transformationdualin} gives back formulas \eqref{transformationdual}, which are written in canonical coordinates for $\circ$. 
 \endproof

Combining the previous Proposition with Theorem \ref{conditionsflatintrinsic} we obtain that the conditions for a density of conservation law $A$ to map a semisimple bi-flat $F$-manifold $(M,\nabla^{(1)},\nabla^{(2)},\circ,*,e,E)$ to another semisimple bi-flat $F$-manifold 
$(M,\tilde{\nabla}^{(1)},\tilde{\nabla}^{(2)},\circ,*,e,E)$ via the map $\psi_A$ are given by the system of equations (here $c^l_{pq}$ are the structure constants of $\circ$, while $c^{*l}_{pq}$ are the structure constants of $*$): 
\beq\label{simultaneousbiflat}
\begin{split}
\nabla^{(1)}_{q}\nabla^{(1)}_{p} A=e(\ln{A})c^l_{pq}\nabla^{(1)}_{l} A,\\
\nabla^{(2)}_{q}\nabla^{(2)}_{p} A=E(\ln{A})c^{*l}_{pq}\nabla^{(2)}_{l} A,
\end{split}
\eeq
where $p$ is possibly equal to $q$.
Before we construct concrete examples of reciprocal transformations preserving the bi-flat $F$-manifold structure, it is convenient to have a simpler form of equations \eqref{simultaneousbiflat}. This is provided by the following:

\begin{proposition}\label{simultaneousbiflat5}
Equations \eqref{simultaneousbiflat} are equivalent to the following system for $A$ the first equation is written in canonical coordinates
 for $\circ$ or for $*$):
\beq\label{simultaneousbiflat2}
\begin{split}
\p_i \p_j A-\Gamma^{i}_{ij}\p_i A-\Gamma^j_{ji}\p_j A=0,\\
e(A)=hA,\\
E(A)=kA.
\end{split}
\eeq
where $h$ and $k$ are constants. 
\end{proposition}
Before giving a proof, let us observe that in the first equation of \eqref{simultaneousbiflat2} it is not necessary to specify if we are considering the first or the second connection, because $\nabla^{(1)}$ and $\nabla^{(2)}$ are almost hydrodynamically equivalent, and thus the Christoffel symbols appearing there are the same for the two connections. 

\proof
For $p\neq q$ both right hand sides of \eqref{simultaneousbiflat} vanish, since $c^{l}_{pq}=\delta^l_p\delta^l_q$ and $c^{*l}_{pq}=\f{1}{u^l}\delta^l_p\delta^l_q$ in canonical coordinates $(u^1, \dots, u^n)$ for $\circ$ (the derivation of the expression for $c^{*l}_{pq}$ can be found in \cite{AL1}). As already observed above, due to the fact that $\nabla^{(1)}$ and $\nabla^{(2)}$ are almost hydrodynamically equivalent, we have that $\nabla^{(1)}_{q} \p_p A=\p_p \p_q A-\Gamma^{p}_{pq}\p_p A-\Gamma^q_{qp}\p_q A=\nabla^{(2)}_{q} \p_p A$ and this gives us the first equation in \eqref{simultaneousbiflat2}. 
For $p=q$, we already know that the first equation in  \eqref{simultaneousbiflat} is equivalent to $e(A)=hA$, $h$ constant, because of Corollary \ref{simplecorollary}. 

Now we focus on the second equation of \eqref{simultaneousbiflat} for $p=q$. In this case we have 
$$\nabla^{(2)}_{p} \p_p A=E(\ln(A))\f{1}{u^p}\p_p A.$$
The left hand side $\nabla^{(2)}_{p} \p_p A$ is equal to 
$$\p^2_pA-\Gamma^{(2)p}_{pp}\p_p A-\sum_{l,\, l\neq p}\Gamma^{(2)l}_{pp}\p_l A.$$
At this point we need the Christoffel symbols for the second connection $\nabla^{(2)}$ appearing in the expression above, given in equations \eqref{dualconnection}.
Substituting the expressions in equations \eqref{dualconnection} into the left hand side of \eqref{simultaneousbiflat} for $p=q$ we obtain 
\beq\label{auxiliary333}\p^2_p A+\left(\sum_{l,\, l\neq p} \f{u^l}{u^p}\Gamma^p_{pl}+\f{1}{u^p} \right)\p_p A+\sum_{l,\, l\neq p}\f{u^l}{u^p}\Gamma^l_{lp}\p_l A=\f{1}{A}E(A)\f{1}{u^p}\p_p A,\eeq
where we dropped the distinction between first and second connections, since for the Christoffel symbols entering in \eqref{auxiliary333} they are equal. 
Since $A$ is a density of conservation law, it satisfies 
$$\p_l \p_p A-\Gamma^l_{lp}\p_l A-\Gamma^p_{pl}\p_p A=0,\quad p\neq l$$
and multiplying by $\f{u^l}{u^p}$ and summing over $l$ for $l\neq p$ one gets
$$\sum_{l,\, l\neq p}\left(  \f{u^l}{u^p}\Gamma^p_{pl}p_p A+\f{u^l}{u^p}\Gamma^l_{lp}\p_l A  \right)=\sum_{l,\, l\neq p}\f{u^l}{u^p}\p_l \p_p A.$$
Substituting this into equation \eqref{auxiliary333} and multiplying both sides by $u^p$ we obtain 
\beq\label{auxiliary444}
u^p \p^2_p A+\sum_{l, \, l\neq p}\p_p(u^l \p_l A)+\p_p A=\f{1}{A}E(A)\p_p A.
\eeq
The left hand side of \eqref{auxiliary444} is readily recognized to be $\p_p(E(A))$ so we are reduced to 
$$\p_p(E(A))=\f{E(A)}{A}\p_p A.$$
Dividing both sides by $A$ we have
$$\f{\p_p(E(A))}{A}-\f{E(A)\p_p A}{A^2}=0,$$
which is just 
$$\p_p\left(\f{E(A)}{A}\right)=0,$$
which is just the last equation of \eqref{simultaneousbiflat2}. 
\endproof
Let us remark that system \eqref{simultaneousbiflat2} does not admit solutions for arbitrary values of the constants $h$ and $k$. 
In particular $h$ must vanish.

\begin{theorem}\label{hiszero}
The system $e(A)=hA,\,E(A)=kA$ admits non trivial
 solutions iff $h=0$.
\end{theorem}

\n
\emph{Proof}. The general solution of the equation
 $e(A)=hA$ can be written in the form:
$$A=e^{hu^1}F(u^2-u^1,\dots,u^n-u^1).$$
Substituting $A=e^{hu^1}F(u^2-u^1,\dots,u^n-u^1)$ into the equation $E(A)=kA$ we obtain
$$e^{hu^1}
\left[\left(\sum_{l=2}^n u^l\d_l+u_1\d_1\right) F(u^2-u^1,\dots,u^n-u^1)-(k-hu^1)F(u^2-u^1,\dots,u^n-u^1)\right]=0.$$
Introducing the variables $\tilde{u}^1=u^1,\,\tilde{u}^l=u^l-u^1,l=2,\dots,n$ and dividing by
 $e^{hu^1}$ the last  equation reduces to
$$\sum_{l=2}^n\tilde{u}^l\d_l F(\tilde{u}^2,\dots,\tilde{u}^n)-(k-h\tilde{u}^1)F(\tilde{u}^2,\dots,\tilde{u}^n)=0,$$
where in the equation above $\p_l:=\f{\p}{\p \tilde u^l}$. 
If $h\ne 0$ differentianting both sides with respect to $\tilde{u}^1$ we obtain 
$$-hF(\tilde{u}^2,\dots,\tilde{u}^n)=0.$$
\endproof

\begin{corollary}\label{aiuto999}
Necessary condition for a density of conservation law $A$ to map a semisimple
 bi-flat $F$-manifold $(M,\nabla^{(1)},\nabla^{(2)},\circ,*,e,E)$ to another semisimple bi-flat $F$-manifold given by
 data $(M,\tilde{\nabla}^{(1)},\tilde{\nabla}^{(2)},\circ,*,e,E)$ is to be a homogeneous flat coordinate  of the natural connection $\nabla^{(1)}$.
\end{corollary}

\n
\emph{Proof}. Due to the Theorem \ref{hiszero}, densities of conservation law preserving the flatness of 
 both the connections $\nabla^{(1)}$ and $\nabla^{(2)}$ necessarily satisfy the system
\beq
\begin{split}
\p_i \p_j A-\Gamma^{i}_{ij}\p_i A-\Gamma^j_{ji}\p_j A=0,\\
e(A)=0.\\
E(A)=kA,
\end{split}
\eeq
for some constant $k$.
On the other hand, homogeneous flat coordinates $f$ of the natural connection are the
 solutions of the system
\beq
\begin{split}
\p_i \p_j f-\Gamma^{i}_{ij}\p_i f-\Gamma^j_{ji}\p_j f=0,\\
\d_i^2 f-\Gamma^l_{ii}\d_l f=\\
\d_i^2 f+\sum_{l\ne i}\left(\Gamma^i_{il}\d_i f+\Gamma^l_{il}\d_l f\right)=\d_i e(f),\\
E(f)=k'f,
\end{split}
\eeq
where the last equality in the second equation is obtained using the first equation and $k'$ is a constant.
Comparing the two systems we obtain the result. Notice that the condition is just necessary since if $f$ is a homogenous flat coordinate
 for the natural connection, then it does not follow in general that $e(f)=0$.
\endproof

Summarizing we have the following Theorem concerning reciprocal transformations preserving the bi-flat $F$-manifold structure:

\begin{theorem}\label{theorembiflat}
Let $(M,\nabla^{(1)},\nabla^{(2)},\circ,*,e,E)$ be a (semisimple) bi-flat $F$-manifold and $A$ any homogeneous
 flat coordinate for $\nabla^{(1)}$ satisfying the further condition $e(A)=0$. Then the sextuple
$(M,\tilde{\nabla}^{(1)},\tilde{\nabla}^{(2)},\circ,*,e,E)$
  with connections $\tilde\nabla^{(1)}$ and $\tilde\nabla^{(1)}$ whose Christoffel symbols are defined respectively by the formulas
\beq\label{tildenabla22}
\tilde\Gamma^{(1)i}_{jk}=\Gamma^{(1)i}_{jk}+c^i_{jk}e^l\omega^{(1)}_l-c^i_{lj}\omega^{(1)}_ke^l
+c^l_{jk}\omega^{(1)}_le^i-c^i_{lk}\omega^{(1)}_je^l, \quad \omega^{(1)}_l:=\nabla^{(1)}_l(\ln( A))
\eeq
and
\beq\label{tildenabla33}
\tilde\Gamma^{(2)i}_{jk}=\Gamma^{(2)i}_{jk}+c^{*i}_{jk}E^l\omega^{(2)}_l-c^{*i}_{lj}\omega^{(2)}_kE^l+c^{*l}_{jk}\omega^{(2)}_lE^i
-c^{*i}_{lk}\omega^{(2)}_jE^l, \quad \omega^{(2)}_l
:=\nabla^{(2)}_l(\ln( A))
\eeq
is a new  (semisimple) bi-flat $F$-manifold.
\end{theorem}


\section{Determination of the orbit structure}
We start with a given semi-Hamiltonian system of hydrodynamic type 
\beq\label{orbit1}u^i_t=v^{(0)i}u^i_x, \quad i=1,\dots, n,\eeq
whose associated natural connection $\nabla^{(0)}$ is flat. 
The goal of this Section is to determine the orbit structure of all semi-Hamiltonian systems of hydrodynamic type that can be obtained starting from \eqref{orbit1}, using reciprocal transformations of the form considered here preserving the flatness of the natural connection. 
Only in this Section the upper index among round brackets refers to the level of iteration: namely $\nabla^{(0)}$ is the starting flat natural connection, $\nabla^{(1)}$ is the one obtained from $\nabla^{(0)}$ using the class of reciprocal transformation under consideration and so on. 

Consider the equation describing density of conservation law for \eqref{orbit1}
\beq\label{orbit2}
\p_i \p_j A-\Gamma^{(0)i}_{ij}\p_i A-\Gamma^{(0)j}_{ji}\p_j A=0, \quad i\neq j
\eeq
without imposing any other conditions on $A$. We denote with $\A^{(0)}$ the vector space over $\mathbb{C}$ of all solutions of \eqref{orbit2}, which is in general infinite dimensional. Indeed, $\A^{(0)}$ can be written via a direct sum decomposition as 
$$\A^{(0)}=\oplus_{h\in \mathbb{C}}\A^{(0)}_h,$$
where  $\A^{(0)}_h$ denotes the finite dimensional vector space of all solutions of the system 
\beq\label{orbit3}
\p_i \p_j A-\Gamma^{(0)i}_{ij}\p_i A-\Gamma^{(0)j}_{ji}\p_j A=0, \quad i\neq j, \quad e(A)=hA.
\eeq

Now we recall some well-known facts. Let us fix $A^{(0)}\in \A^{(0)}$ and consider the reciprocal transformation $\psi_{A^{(0)}}$ generated by this density of conservation law. Unless $A^{(0)}\in \A^{(0)}_h$ for some $h$, we know that the new natural connection $\nabla^{(1)}$ associated to functions 
$\Gamma^{(1)i}_{ij}-\p_j \ln(A^{(0)})$ will not be flat, nevertheless, since reciprocal transformations preserve the semi-Hamiltonian property, it makes sense to consider the vector space $\A^{(1)}$ of all solutions of 
\beq\label{orbit4}\p_i \p_j A-\Gamma^{(1)i}_{ij}\p_i A-\Gamma^{(1)j}_{ji}\p_j A=0, \quad i\neq j.\eeq It is well-known that any element of  $A^{(1)\alpha}\in \A^{(1)}$ can be written as
$$A^{(1)\alpha}=\f{A^{(0)\alpha}}{A^{(0)}},$$
where $A^{(0)\alpha}$ varies in $\A^{(0)}$.
In other terms, each solution of \eqref{orbit4} can be obtained from a corresponding solution of \eqref{orbit2} dividing it by $A^{(0)}$, the density of conservation law generating the reciprocal transformation that links $\nabla^{(0)}$ to $\nabla^{(1)}$. 
Suppose now we fix $A^{(1)}\in \A^{(1)}$ and consider the reciprocal transformation generated by it, sending $\nabla^{(1)}$ to $\nabla^{(2)}$. Any element in $\A^{(2)}$, the vector space of all solutions of 
$$\p_i \p_j A-\Gamma^{(2)i}_{ij}\p_i A-\Gamma^{(2)j}_{ji}\p_j A=0, \quad i\neq j,$$
can be obtained as $A^{(2)\beta}=\f{A^{(1)\beta}}{A^{(1)}}$, where $A^{(1)\beta}$ varies in $\A^{(1)}$. But using the fact that $A^{(1)\beta}=\f{A^{(0)\beta}}{A^{(0)}}$ and the fixed element  $A^{(1)}$ can be written as $\f{\bar A^{(0)}}{A^{(0)}}$ for a suitable ${\bar A^{(0)}}$ we have that any element of $\A^{(2)}$ can be written as 
$$A^{(2)\beta}=\f{A^{(1)\beta}}{A^{(1)}}=\f{\f{A^{(0)\beta}}{A^{(0)}}}{\f{\bar A^{(0)}}{A^{(0)}}}=\f{A^{(0)\beta}}{\bar A^{(0)}}.$$
This shows the well-know property that $\nabla^{(2)}$ can be obtained directly from $\nabla^{(0)}$ using the reciprocal transformation generated by $\bar A^{(0)}.$ Iterating this argument, we arrive at the following characterization of the orbit structure, which is well-known. The semi-Hamiltonian system of hydrodynamic type that can be reached starting from $u^i_t=v^{(0)i}u^i_x$ via any sequence of reciprocal transformations of the form \eqref{RT} are given exactly by the systems of the form 
\beq\label{orbit10}u^i_t=\left( A^{(0)}v^{(0)i}-B^{(0)}\right)u^i_x,\eeq
where $A^{(0)}$ varies in  $\A^{(0)}$ and $B^{(0)}$ is the corresponding conserved current. 
 
Now we apply these observations to the case in which we require the flatness of the sequence of natural connections $\nabla^{(1)}, \nabla^{(2)}, \dots, \nabla^{(k)}, \dots$ obtained starting from $\nabla^{(0)}$. 
Since flatness has to be preserved, it is clear that at each step, the density of conservation law $A^{(k)}\in \A^{(k)}$ used do define $\psi_{A^{(k)}}$ sending $\nabla^{(k)}$ to $\nabla^{(k+1)}$ has to satisfy the condition $e(A^{(k)})=hA^{(k)}$, for some $h\in \mathbb{C}$  (where $h$ possibly depends on the iteration level $k$), because of Corollary \ref{simplecorollary}. Therefore $A^{(k)}\in \A^{(k)}_h$ for some $h$. 

Combining these remarks we obtain the following
\begin{proposition}
Consider the semi-Hamiltonian system of hydrodynamic type $u^i_t=v^{(0)i}u^i_x$ and assume that the associated natural connection $\nabla^{(0)}$ is flat. Then the semi-Hamiltonian systems that can be obtained starting from $u^i_t=v^{(0)i}u^i_x$ through iteration of any sequence of reciprocal transformations of the form \eqref{RT} preserving the flat $F$-manifold structure are exactly those that can be written as
\beq\label{orbit11}u^i_t=\left( A^{(0)}_hv^{(0)i}-B^{(0)}_h\right)u^i_x, \eeq
where $A^{(0)}_h$ varies in $\A^{(0)}_h$, $h$ varies in $\mathbb{C}$ and $B^{(0)}_h$ is the conserved current associated to $A^{(0)}_h$.
\end{proposition}
\proof Fix $A^{(0)}_h \in \A^{(0)}_h$ for some $h$ and consider the map $\psi_{A^{(0)}_h}$. As recalled above this map gives a bijection from $\A^{(0)}$ to $A^{(1)}$. More precisely, this map gives a bijection 
$$\psi_{A^{(0)}_h}: \A^{(0)}_k \rightarrow \A^{(1)}_{k-h}, \quad \forall k \in \mathbb{C},$$
since for each $A\in \A^{(0)}_k$ it is immediate to check that $e\left( \f{A}{A^{(0)}_h}\right)=(k-h)\left(  \f{A}{A^{(0)}_h}\right)$.
Therefore iteration of maps of the form 
$$\nabla^{(0)}\stackrel{\psi_{A^{(0)}_{h^0}}}{\rightarrow}\nabla^{(1)}\stackrel{\psi_{A^{(1)}_{h^1}}}{\rightarrow}\nabla^{(2)}\rightarrow \dots \nabla^{(m-1)}\stackrel{\psi_{A^{(m-1)}_{h^m}}}{\rightarrow}\nabla^{(m)},$$
preserves the flatness of all natural connections and it is compatible with the direct sum decompositions of all the vector spaces $\A^{(0)}$, $\A^{(1)}, \dots, \A^{(m)}.$
By what we recalled above the compositions of these maps can be still be expressed directly as a map from $\nabla^{(0)}$ to $\nabla^{(m)}$ for a suitable density of conservation law $A^{(0)}_l$ for some suitable $l$. 
\endproof

\section{Reciprocal transformations and the Darboux-Egorov system}
In \cite{AL2}, we showed how to construct bi-flat $F$-manifold structures starting from solutions of a suitably augmented Darboux-Egorov system, aDE system for short. In this section we explore the action of the reciprocal transformations introduced before on the solutions of the aDE system.
We recall the main set-up from \cite{AL2}. Consider the following system 
\beq\label{darbouxegorov}
\begin{split}
\d_k\beta_{ij}=\beta_{ik}\beta_{kj},\quad k\ne i\ne j\ne k,\\
e(\beta_{ij})=0, \quad i\neq j\\
E(\beta_{ij})=-\beta_{ij}, \quad i\neq j
\end{split}
\eeq
where $e=\sum\f{\d}{\d u^i}$, $E=\sum_i u^i\f{\d}{\d u^i}$ for unknown functions $\beta_{ij}(u^1, \dots, u^n)$, $i\neq j$. 
The functions $\beta_{ij}$ are usually known as Ricci's rotation coefficients in the literature. 
 Suppose a solution $\beta_{ij}$ ($i\neq j$, $i,j=1, \dots, n$) is known and consider the following system for the unknown functions $H_i(u^1, \dots, u^n)$, $i=1,\dots, n$ (called Lam\'e coefficients):
\beq\label{lame}
\begin{split}
\p_j H_i=\beta_{ij}H_j, \quad i\neq j,\\
e(H_i)=0,\\
E(H_i)=-d H_i,
\end{split}
\eeq
for a suitable constant $d$. It turns out that it is possible to choose suitable constants $d$ such that the system \eqref{lame} admits nontrivial solutions (see \cite{AL2}). Then if a solution $H_i$ ($i=1, \dots, n$) of \eqref{lame} is known corresponding to a known solution $\beta_{ij}$ ($i\neq j$, $i,j=1, \dots, n$), it is possible to define functions $\Gamma^{i}_{ij}:=\f{H_j}{H_i}\beta_{ij}$, $i\neq j$ and via them a corresponding natural connection $\nabla^{(1)}$, using formulas \eqref{nat-conn-eps}, and a dual connection $\nabla^{(2)}$, using formulas \eqref{dualconnection}.
Notice also that, by construction, $\Gamma^{(1)i}_{ij}=\Gamma^{(2)i}_{ij}$, $i\neq j$.  

It has been proved in \cite{AL2} that this construction gives rise to a bi-flat $F$-manifold structure $(M, e, E, \nabla^{(1)}, \nabla^{(2)}, \circ, *)$, with $e$ and $E$ given as above and $c^i_{jk}=\delta^i_j\delta^i_k$ and $c^{*i}_{jk}=\f{1}{u^i}\delta^i_j \delta^i_k$. This structure is associated to any solution $(\beta_{ij}, H_i)$ of \eqref{darbouxegorov} and \eqref{lame} respectively. It is therefore natural to investigate how the reciprocal transformations studied above acts on solutions of  \eqref{darbouxegorov} and \eqref{lame}. We proceed as follows. 

We know that $\tilde \Gamma^i_{ij}=\Gamma^i_{ij}-\p_j \ln(A)$, $i\neq j$ and we can write it as
$$\tilde \Gamma^i_{ij}=\f{H_j}{H_i}\beta_{ij}-\f{H_j}{H_i}\left(\f{H_i}{H_j}\p_j \ln(A)\right)=\f{\tilde H_j}{\tilde H_i}\tilde \beta_{ij},\quad i\neq j$$
where 
\beq\label{newbeta}
\tilde \beta_{ij}:=\beta_{ij}-\f{H_i}{H_j}\p_j \ln(A), \quad i\neq j,
\eeq
and 
\beq\label{newh}
\tilde H_{i}:=\f{H_i}{A}.
\eeq
The reason for choosing to define $\tilde H_i$ as in \eqref{newh} stems from the fact that in this way the equation 
$$\p_j \tilde H_i=\tilde \beta_{ij}\tilde H_j, \quad i\neq j,$$
namely the first of \eqref{lame} is identically satisfied: 
$$\tilde \beta_{ij}=\f{\p_j \tilde H_i}{\tilde H_j}=\f{\p_j H_i}{H_j}-A\f{H_i}{H_j}\f{\p_j A}{A^2}=\beta_{ij}-\f{H_i}{H_j}\p_j \ln(A), \quad i\neq j$$
where in the last equality we used the first equation of \eqref{lame} for $\beta_{ij}$, $H_i$. 
Formulas \eqref{newbeta} and \eqref{newh} represents the action of the reciprocal transformations studied in the previous Sections at the level of rotation and Lam\'e coefficients, respectively. 

We have the following result, which might be already known in the classical literature, maybe expressed using a different language. 
\begin{theorem}\label{reciprocaldarboux}
Suppose $\beta_{ij}$ satisfies system \eqref{darbouxegorov} and $H_i$ satisfies system \eqref{lame} for the corresponding $\beta_{ij}$ for suitable $d$. Assume that $A$ satisfies the following system
$$\p_j\p_i A-\Gamma^i_{ij}\p_i A-\Gamma^j_{ji}\p_j A=0,\; i\neq j\quad e(A)=0, \quad E(A)=kA$$
for $\Gamma^i_{ij}:=\f{H_j}{H_i}\beta_{ij}$ and for some constant $k$.
Then 
$$\tilde \beta_{ij}:=\beta_{ij}-\f{H_i}{H_j}\p_j \ln(A), \quad i\neq j,$$
$$\tilde H_{i}:=\f{H_i}{A},$$
satisfy systems  \eqref{darbouxegorov} and \eqref{lame} respectively, with $d$ replaced by $d+k$ in the last equation of system \eqref{lame}. 
\end{theorem}
Before giving a proof, let us remark that the conditions imposed on $A$ in Theorem \ref{reciprocaldarboux} are exactly those that ensure that the reciprocal transformation generated by $A$ preserves the bi-flat $F$-manifold structure.

\proof
We assume till the end of the proof that all distinct indices are different and repeated indices are not summed. 
We first analyze system \eqref{lame}. By the definition of $\tilde H_i$ we have
$$\f{\p_j \tilde H_i}{\tilde H_j}=\f{\p_j H_i}{H_j}-\f{H_i}{H_j}\p_j \ln(A)=\tilde \beta_{ij}, \quad i\neq j$$
where the last equality is due to the definition of $\tilde \beta_{ij}$. Therefore the first of \eqref{lame} is automatically fulfilled. 
Moreover $$e(\tilde H_i)=\f{e(H_i)}{A}-\f{H_i}{A^2}e(A)=0,$$
because $e(H_i)=0$ and $e(A)=0$ by assumption. Finally, 
$$E(\tilde H_i)=\f{E(H_i)}{A}-\f{H_i}{A^2}E(A)=-(d+k)\tilde H_i,$$
since $E(H_i)=-dH_i$ and $E(A)=kA$. 

Now we consider system \eqref{darbouxegorov}. We have 
$$e(\tilde \beta_{ij})=e\left(\beta_{ij}-\f{H_i}{H_j}\f{\p_j A}{A} \right)=e(\beta_{ij})-\f{e(H_i)}{H_j}\f{\p_j A}{A}+\f{H_i}{H_j^2}e(H_j)\f{\p_j A}{A}$$
$$-\f{H_i}{H_j}\f{e(\p_jA)}{A}+\f{H_i}{H_j}\f{\p_j A}{A^2}e(A)=0,\quad i\neq j$$
since $e(H_i)=0, e(A)=0$ and $e(\p_j A)=\p_j (e(A))=0$.
Analogously we have 
$$E(\tilde \beta_{ij})=E\left(\beta_{ij}-\f{H_i}{H_j}\f{\p_j A}{A} \right)=E(\beta_{ij})-\f{E(H_i)}{H_j}\f{\p_j A}{A}+\f{H_i}{H_j^2}E(H_j)\f{\p_j A}{A}$$
$$-\f{H_i}{H_j}\f{E(\p_jA)}{A}+\f{H_i}{H_j}\f{\p_j A}{A^2}E(A), \quad i\neq j.$$
Since $E(\beta_{ij})=-\beta_{ij}$, $E(A)=kA$, $E(\p_j A)=(k-1)\p_j A$, $E(H_i)=-dH_i$, we have that $E(\tilde \beta_{ij})$ is equal to  
$$-\beta_{ij}+d\f{H_i}{H_j}\f{\p_j A}{A}-d\f{H_i}{H_j}\f{\p_j A}{A}-(k-1)\f{H_i}{H_j}\f{\p_j A}{A}+k\f{H_i}{H_j}\f{\p_j A}{A}=-\beta_{ij}+\f{H_i}{H_j}\f{\p_j A}{A}=-\tilde \beta_{ij}.$$
To conclude the proof, it remains to show that also the first equation of \eqref{darbouxegorov} is satisfied, namely we have to prove that 
\beq\label{proofdarbouxegorov}
\p_k \tilde\beta_{ij}=\tilde \beta_{ik}\tilde\beta_{kj}, \quad i\neq j\neq k\neq i. 
\eeq
Substituting $\tilde\beta_{ij}=\beta_{ij}-\f{H_i}{H_j}\f{\p_j A}{A}$ in \eqref{proofdarbouxegorov} we find
\beq\label{proofdarboux2}
\p_k\left(\beta_{ij}- \f{H_i}{H_j}\f{\p_j A}{A}\right)=\left(\beta_{ik}-\f{H_i}{H_k}\f{\p_k A}{A} \right)\left(\beta_{kj}- \f{H_k}{H_j}\f{\p_j A}{A} \right).
\eeq 
Expanding both sides of \eqref{proofdarboux2}, using the fact that $\beta_{ij}$ indeed satisfies the first of \eqref{darbouxegorov} and cancelling two terms on both sides that are obviously equal we are left to prove that 
\beq\label{proofdarboux3}
\begin{split}
-\f{\p_k H_i}{H_j}\f{\p_j A}{A}+\f{H_i}{H_j^2}\p_k H_j\f{\p_j A}{A}-\f{H_i}{H_j}\f{\p_k \p_j A}{A}=-\f{H_i}{H_k}\f{\p_k A}{A}\beta_{kj}-\f{H_k}{H_j}\f{\p_j A}{A}\beta_{ik}.
\end{split}
\eeq
Since 
$$\p_k \p_j A=\Gamma^k_{kj}\p_k A+\Gamma^j_{jk}\p_j A=\f{H_j}{H_k}\beta_{kj}\p_k A+\f{H_k}{H_j}\beta_{jk}\p_j A,$$
substituting this expression in the left hand side of \eqref{proofdarboux3} and using $\p_k H_i=H_k \beta_{ik}$ ($i\neq k$) and $\p_k H_j=H_k \beta_{jk}$ ($j\neq k$) it is immediate to see that \eqref{proofdarboux3} is identically satisfied and therefore \eqref{proofdarbouxegorov} is satisfied too. 
\endproof

The meaning of Theorem \ref{reciprocaldarboux} is that via the action of the reciprocal transformations we are considering, if we start from a bi-flat $F$-manifold structure arising via the augmented Darboux-Egorov system, we still obtain a bi-flat $F$-manifold structure associated to the augmented Darboux-Egorov system. 

\section{The $\epsilon$-system}
In this section we recall few facts about the $\epsilon$-system, since the examples we construct in the next few Sections are based on it. 
The semi-Hamiltonian system 
$$u^i_t=\left(u^i-\epsilon\sum_{k=1}^n u^k \right)u^i_x, \quad i=1, \dots, n,$$
is known in the literature \cite{Pv} as the $\epsilon$-system. 
In \cite{LP} it has been proved that the natural connection associated to the $\epsilon$-system 
\begin{eqnarray*}
\Gamma^{(1)i}_{jk}&=&0,\qquad\forall i\ne j\ne k \ne i,\\
\Gamma^{(1)i}_{jj}&=&-\Gamma^{(1)i}_{ij},\qquad i\ne j,\\
\Gamma^{(1)i}_{ij}&=&\f{\epsilon}{u^i-u^j},\qquad i\ne j,\\
\Gamma^{(1)i}_{ii}&=&-\sum_{l\ne i}\Gamma^{(1)i}_{li}.
\end{eqnarray*}
is flat. Moreover, it has been observed in \cite{AL2} that to the $\epsilon$-system one can associate a second flat connection $\nabla^{(2)}$ 
defined by
\begin{eqnarray*}
\Gamma^{(2)i}_{jk}&=&0,\qquad\forall i\ne j\ne k \ne i,\\
\Gamma^{(2)i}_{jj}&=&-\f{u^i}{u^j}\Gamma^{(2)i}_{ij},\qquad i\ne j,\\
\Gamma^{(2)i}_{ij}&=&\f{\epsilon}{u^i-u^j},\qquad i\ne j,\\
\Gamma^{(2)i}_{ii}&=&-\sum_{l\ne i}\f{u^l}{u^i}\Gamma^{(2)i}_{li}-\f{1}{u^i},
\end{eqnarray*}
 such that the product $c^i_{jk}=\delta^i_j\delta^i_k$ and $c^{*i}_{jk}=\frac{1}{u^i}\delta^i_j\delta^i_k$, $e=\sum_{k=1}^n \partial_k$ and $E=\sum_{k=1}^n u^k\partial_k$ define
 a bi-flat semisimple $F$-manifold structure.
 
 In Sections \ref{examplesn2}, \ref{examplesn3} we are going to construct reciprocal transformations sending $\nabla^{(1)}$ to another flat connection, but in general not preserving the flatness of $\nabla^{(2)}$, in dimensions $2$ and $3$ respectively. Instead, in Section \ref{examplesbiflat} we consider reciprocal transformations preserving the bi-flat $F$-manifold structure, namely sending both connections $\nabla^{(1)}$ and $\nabla^{(2)}$ to different flat connections $\tilde \nabla^{(1)}$, $\tilde \nabla^{(2)}$.  
 
\section{Examples in dimension two}\label{examplesn2}

In this section we work out an example in dimension two. Let $A(u^1, u^2)$ a density of conservation law associated to the $\epsilon$-system. Then it satisfies the single PDE
\beq\label{densityn2}
\p_1 \p_2 A-\Gamma^1_{12}\p_1 A-\Gamma^2_{21}\p_2 A=0, 
\eeq
where $\Gamma^1_{12}:=\frac{\epsilon}{u^1-u^2}$, and $\Gamma^{2}_{21}:=\frac{\epsilon}{u^2-u^1}.$
We also impose the condition that allows to preserve the flatness of the natural connection $\nabla^{(1)}$, namely 
\beq\label{densityn22}e(A)=h A,\eeq
for some parameter $h$, and we consider the case $h\neq0$. The general solution of \eqref{densityn22} can be easily seen to be expressible as 
\beq\label{generalsoln2}A(u^1, u^2):=f(u^2-u^1){{\rm e}^{hu^1}},\eeq
where $f$ is an arbitrary smooth function of a single variable. Substituting the general solution \eqref{generalsoln2} into the single PDE for density of conservation law \eqref{densityn2}, after setting $z:=u^2-u^1$ and after some straightforward computations we obtain:
\beq\label{densityn23}
z\f{d^2f(z)}{dz^2}-zh\f{d f(z)}{dz}+2\epsilon\f{d f(z)}{dz}-\epsilon f(z)h=0.
\eeq
The general solution of \eqref{densityn23} is given by (for $h\neq 0$)
\beq\begin{split}f \left( z \right) ={c_1}\,{{\rm e}^{\frac{1}{2}\,zh}}\,{z}^{\frac{1}{2}-\epsilon}\,
{{\rm J}\left(\epsilon-\frac{1}{2},\,\frac{1}{2}\,\sqrt {-{h}^{2}}z\right)}\\
+{c_2}\,{
{\rm e}^{\frac{1}{2}\,zh}}\,{z}^{\frac{1}{2}-\epsilon}\,
{{\rm Y}\left(\epsilon-\frac{1}{2},\,\frac{1}{2}\,\sqrt {-{h}^{2}}z\right)},\end{split}\eeq
where $c_1, c_2$ are constants of integration; moreover ${\rm J}$ and ${\rm Y}$ are Bessel functions of the first and second kind, respectively, with parameter $\epsilon-\frac{1}{2}$ and with indeterminate given by $\frac{1}{2}\sqrt {-{h}^{2}}z$ in both cases. 

Therefore the general solution $A(u^1, u^2)$ of the system (\ref{densityn2}, \, \ref{densityn22}), for $h\neq 0$ is given by 
\beq\label{gensoln2}
\begin{split}
A(u^1, u^2)={{\rm e}^{hu^1}}\,{c_1}\,{{\rm e}^{\frac{1}{2}\,(u^2-u^1)h}}\,{(u^2-u^1)}^{\frac{1}{2}-\epsilon}\,
{{\rm J}\left(\epsilon-\frac{1}{2},\,\frac{1}{2}\,\sqrt {-{h}^{2}}(u^2-u^1)\right)}\\
+{{\rm e}^{hu^1}}{c_2}\,{
{\rm e}^{\frac{1}{2}\,(u^2-u^1)h}}\,{(u^2-u^1)}^{\frac{1}{2}-\epsilon}\,
{{\rm Y}\left(\epsilon-\frac{1}{2},\,\frac{1}{2}\,\sqrt {-{h}^{2}}(u^2-u^1)\right)}.
\end{split}
\eeq

To determine the current $B$ associated to the density of conservation laws $A$, we impose the condition $d(Adx+B dt)=0$ which entails $A_t=B_x$. In this case, it gives rise to the two PDEs:
\beq\label{current1}\p_1 B-v^1\p_1 A =0, \quad \p_2 B-v^2\p_2 A=0,\eeq
where $v^i$ is the $i$-th characteristic velocity, namely $u^i-\epsilon(u^1+u^2)$ for the two components $\epsilon$-system. 
For general $\epsilon$ it is  computationally difficult to find the general solution of system \eqref{current1}.
We present results for $\epsilon=\pm 1$. 
\subsection{Case $\epsilon=1$ and $h\neq 0$ }
In this case we have 
\beq\label{density111}
A(u^1, u^2)=c_1\f{{\rm e}^{hu^1}}{u^2-u^1}+c_2\f{{\rm e}^{hu^2}}{u^2-u^1},
\eeq
and for the corresponding current
\beq\label{density222}
B(u^1, u^2)=c_1\f{{\rm e}^{hu^1}u^2}{u^1-u^2}+c_2\f{{\rm e}^{hu^2}u^1}{u^1-u^2}+c_3,
\eeq
where $c_1, c_2, c_3$ are arbitrary constants. The $\epsilon$ system $u^i_t=v^i u^i_x$ it thus mapped to $u^i_{\tilde t}=[Av^i-B]u^i_{\tilde x}$ with $A$ and $B$ given in \eqref{density111} and \eqref{density222} respectively. 

\subsection{Case $\epsilon=1$ and  $h=0$}
In this case we have 
\beq\label{density121212}
A(u^1, u^2)=c_1+c_2\f{1}{u^2-u^1},
\eeq
and for the corresponding current
\beq\label{density222222}
B(u^1, u^2)=\f{u^2c_2}{u^1-u^2}+c_3
\eeq
where $c_1, c_2, c_3$ are arbitrary constants. The $\epsilon$ system $u^i_t=v^i u^i_x$ it thus mapped to $u^i_{\tilde t}=[Av^i-B]u^i_{\tilde x}$ with $A$ and $B$ given in \eqref{density121212} and \eqref{density222222} respectively.

\subsection{Case $\epsilon=-1$ and  $h\neq 0$}
In this case we have 
\beq\label{density333}
A(u^1, u^2)=c_1{\rm e}^{hu^1}\left[hu^2-hu^1+2\right]+c_2{\rm e}^{hu^2}\left[hu^2-hu^1-2\right],
\eeq
and for the corresponding current 
\beq\label{density444}
\begin{split}
B(u^1, u^2)=c_1{\rm e}^{hu^1}\left[6u^1-(2u^1+u^2)(u^1-u^2)h-\f{6}{h} \right]\\
+c_2{\rm e}^{hu^2}\left[-6u^2-(2u^2+u^1)(u^1-u^2)h+\f{6}{h}\right]+c_3,
\end{split}
\eeq
where $c_1, c_2, c_3$ are arbitrary constants. The $\epsilon$ system $u^i_t=v^i u^i_x$ it thus mapped to $u^i_{\tilde t}=[Av^i-B]u^i_{\tilde x}$ with $A$ and $B$ given in \eqref{density333} and \eqref{density444} respectively. 

\subsection{Case $\epsilon=-1$ and  $h=0$}
In this case we have 
\beq\label{density3332}
A(u^1, u^2)=c_1+c_2(u^2-u^1)^3,
\eeq
and for the corresponding current 
\beq\label{density4442}
\begin{split}
B(u^1, u^2)=c_2\f{3}{2}(u^1+u^2)(u^2-u^1)^3+c_3
\end{split}
\eeq
where $c_1, c_2, c_3$ are arbitrary constants.

\section{Examples in dimension three}\label{examplesn3}
In dimension three, we first analyze the case in which only one connection, namely $\nabla^{(1)}$ is preserved. We won't be able to obtain a close formula for the density $A(u^1, u^2, u^3)$ depending on $\epsilon$, but we will present explicit formulas for $\epsilon=\pm 1$ and explain the procedure in general. 

Again we consider the system for the unknown function $A(u^1, u^2, u^3)$ given by 
\beq\label{fundsysn3}
\begin{split}
\f{\p A}{\p u^1}+\f{\p A}{\p u^2}+\f{\p A}{\p u^3}=hA,\\
\f{\p^{2} A}{\p u^2\p u^1} -\f{\e}{u^1-u^2}\f{\p A}{\p u^1}-\f{\e}{u^2-u^1}\f{\p A}{\p u^2}=0,\\
\f{\p^{2} A}{\p u^3\p u^2} -\f{\e}{u^2-u^3}\f{\p A}{\p u^2}-\f{\e}{u^3-u^2}\f{\p A}{\p u^3}=0,\\
\f{\p^{2} A}{\p u^3\p u^1} -\f{\e}{u^3-u^1}\f{\p A}{\p u^3}-\f{\e}{u^1-u^3}\f{\p A}{\p u^1}=0.\\
\end{split}
\eeq

First of all, observe that the general solution of the first equation of the system provides $A(u^1, u^2, u^3)$  factorized as follows: 
\beq\label{recoverA}A(u^1, u^2, u^3)=F(u^2-u^1, u^3-u^2){\rm e}^{h u^2},\eeq
where $F$ is an arbitrary smooth function of two variables. This suggests to rewrite the remaining three equations of system \eqref{fundsysn3} with respect to new variables $x^{22}:=u^2$, $x^{32}:=u^3-u^2$ and $x^{21}=u^2-u^1$ and substitute the expression for $A=F(x^{21}, x^{32}){\rm e}^{h x^{22}}.$

Proceeding in this way and dividing the resulting expressions by the never vanishing term ${\rm e}^{h x^{22}}$, we obtain the system of PDEs for $F(x^{21}, x^{32})$ given by 
\beq\label{fund2sysn3}
\begin{split}
-hx^{21}\f{\p F}{\p x^{21}}+x^{21}\f{\p^2 F}{\p x^{32}\p x^{21}}-x^{21}\f{\p^2 F}{(\p x^{21})^2}-2\e\f{\p F}{\p x^{21}}-\e h F+\e\f{\p F}{\p x^{32}}=0,\\
hx^{32}\f{\p F}{\p x^{32}}-x^{32}\f{\p^2 F}{(\p x^{32})^2}+x^{32}\f{\p^2 F}{\p x^{32} \p x^{21}}+\e h F-2\e \f{\p F}{\p x^{32}}+\e \f{\p F}{\p x^{21}}=0,\\
x^{32}\f{\p^2 F}{\p x^{32} \p x^{21}}+x^{21}\f{\p^2 F}{\p x^{32} \p x^{21}}+\e \f{\p F}{\p x^{32}}+\e \f{\p F}{\p x^{21}}=0.
\end{split}
\eeq

Now we show how system \eqref{fund2sysn3} can be reduced to a single ODE with respect to the variable $x^{32}$ for the function $F$. From the third equation in system \eqref{fund2sysn3}, we express  $\f{\p^2 F}{\p x^{32} \p x^{21}}$ in terms of the other derivatives. Substituting this expression in the first two equations of \eqref{fund2sysn3} we obtain the following two PDEs:
\beq\label{auxiliaryder1}
\begin{split}
hx^{21}x^{32}\f{\p F}{\p x^{21}}+h(x^{21})^2\f{\p F}{\p x^{21}}+3\e x^{21}\f{\p F}{\p x^{21}}+x^{21}x^{32}\f{\p^2 F}{(\p x^{21})^2}\\+(x^{21})^2\f{\p^2 F}{(\p x^{21})^2}
+2\e x^{32}\f{\p F}{\p x^{21}}+\e h F(x^{32}+x^{21})-\e x^{32}\f{\p F}{\p x^{32}}=0,\\ 
h(x^{32})^2\f{\p F}{\p x^{32}}+hx^{32}x^{21}\f{\p F}{\p x^{32}}-(x^{32})^2\f{\p^2 F}{(\p x^{32})^2}-x^{32}x^{21}\f{\p^2 F}{(\p x^{32})^2}\\-3\e x^{32}\f{\p F}{\p x^{32}}+\e h F(x^{32}+x^{21})-2\e x^{21}\f{\p F}{\p x^{32}}+\e x^{21}\f{\p F}{\p x^{21}}=0.
\end{split}
\eeq
Using the second equation in \eqref{auxiliaryder1} to obtain $\f{\p F}{\p x^{21}}$ in terms of the derivatives with respect to $x^{32}$ and substituting back in the first equation of \eqref{auxiliaryder1} we obtain the following fourth order ODE in $x^{32}$:
\beq\label{fundODEn3}
\begin{split}
-\{(x^{32})^2(x^{21}+x^{32})\}\f{d^4 F}{(d x^{32})^4}+\\\{2x^{32}\left(h(x^{32})^2+hx^{21}x^{32}-3\e x^{32}-2x^{32}-x^{21}-2\e x^{21}\right)\}\f{d^3 F}{(d x^{32})^3}\\
+\{ hx^{21}(x^{32}(3+5\e)-4\e^2-h^2  (x^{32})^2-2\e)-11\e x^{32}(1+\e)\\-h^2 (x^{32})^3+(x^{32})^2(6h +8 h \e)-2x^{32} \}\f{d^2 F}{(d x^{32})^2}\\
-\{(\e+1)\left(6\e^2-2\e h x^{21}-8\e h x^{32}+2h^2(x^{32})^2+h^2 x^{21}x^{32}-2hx^{32} \right)\}\f{d F}{d x^{32}}\\+\e h (\e+1)(2\e-h x^{32})F=0,
\end{split}
\eeq
where $F$ is a function of $x^{32}$ and $x^{21}$, but $x^{21}$ appears in the previous equation as a parameter. The general solution of \eqref{fundODEn3} can be expressed as a linear combination of four functions of $x^{32}$, possibly containing $x^{21}$ as a parameter; the coefficients of this linear combination are four functions of $x^{21}$ only. In general, even a symbolic computation program like ${\tt Maple}^{\tt TM}$ is unable to compute the general solution if the parameter $\e$ is left unspecified. 
To construct explicitly solutions to system \eqref{fundsysn3} we consider separately two cases. 

\subsection{Case $\e=1$}
Using ${\tt Maple}^{\tt TM}$, in the case $\epsilon=1$, the general solution to ODE \eqref{fundODEn3} can be written as 
\beq\begin{split}F(x^{21}, x^{32})=\f{f_1(x^{21})}{x^{32}+x^{21}}+\f{f_2(x^{21})}{x^{32}(x^{32}+x^{21})}\\+\f{f_3(x^{21}){\rm e}^{hx^{32}}}{x^{32}(x^{32}+x^{21})}
+\f{f_4(x^{21})\left({\rm e}^{h x^{32}}+h x^{32} {\rm Ei}(1, -hx^{32}) \right)}{x^{32}(x^{32}+x^{21})},\end{split}\eeq
where $f_1(x^{21}), \dots, f_4(x^{21})$ are arbitrary smooth functions of a single variable while ${\rm Ei}(a, z)$ is defined as
$${\rm Ei}(a, z):=\int_1^{+\infty}{\rm e}^{-\lambda z}\lambda^{-a}\, d\lambda, \quad \text{ for } \Re(z)>0,$$
and extended via analytic continuation to the entire complex plane (except $z=0$) with the following expression:
$${\rm Ei}(a, z)=z^{a-1}\Gamma(1-a, z).$$
Here $\Gamma(b, z)$ is the incomplete Gamma function. 
To determine the functions $f_1, \dots, f_4$ appearing in the expression above for $F$ it is necessary to substitute back that expression into system \eqref{fund2sysn3} and perform several computations. We just report the result which is
\beq\label{casee1}
F(x^{21}, x^{32})=c_0\f{1}{x^{32}x^{21}}+c_1\f{{\rm e}^{hx^{32}}}{x^{32}(x^{32}+x^{21})}+c_2\f{{\rm e}^{-hx^{21}}}{x^{21}(x^{32}+x^{21})},
\eeq
where $c_0, c_1, c_2$ are arbitrary constants. Thus formula \eqref{casee1} provides a three parameter family of solutions of \eqref{fund2sysn3}, for $h\neq 0$.  The reason for which \eqref{casee1} provides a three parameter family of solutions for $h\neq 0$ is because of this. For $h\neq 0$, a basis of solutions of \eqref{fundODEn3} is given by $\f{1}{x^{32}+x^{21}},$  $\f{1}{x^{32}(x^{32}+x^{21})},$ 
$\f{{\rm e}^{hx^{32}}}{x^{32}(x^{32}+x^{21})}$, $\f{{\rm e}^{hx^{32}}+hx^{32} {\rm Ei}(1, -hx^{32})}{x^{32}(x^{32}+x^{21})}$, while for $h=0$, a basis of solutions is given by $\f{1}{x^{32}+x^{21}}$, $\f{x^{32}}{x^{32}+x^{21}}$, $\f{1}{x^{32}(x^{32}+x^{21})}$, $\f{\ln(x^{32})}{x^{32}+x^{21}}$.

The corresponding $A$ can be recovered from \eqref{casee1} using \eqref{recoverA}. 
In the particular case $\epsilon=1$, $h=0$, it is possible to find directly the general solution of system \eqref{fund2sysn3}. In this particular case, a three parameter family of solutions determined using ${\tt Maple^{TM}}$ is given by:
$$F(x^{21}, x^{32})=\f{c_1}{x^{32}(x^{21}+x^{32})}+\f{c_1}{x^{21}(x^{32}+x^{21})}+c_2,$$
where $c_0, c_1, c_2$ are arbitrary constants.

\subsection{Case $\e=-1$}
Using ${\tt Maple}^{\tt TM}$, in the case $\epsilon=-1$, the general solution to ODE \eqref{fundODEn3} can be written as (for $h\neq 0$)
\beq\begin{split}F(x^{21}, x^{32})=f_1(x^{21})+f_2(x^{21})x^{32}+f_3(x^{21}){\rm e}^{hx^{32}}(-2+h^2 (x^{32})^2)\\+f_4(x^{21}){\rm e}^{h x^{32}}(x^{32}-h(x^{32})^2),\end{split}\eeq
where as before $f_1, \dots, f_4$ are arbitrary smooth functions of a single variable. The four functions $1$, $x^{32}$, ${\rm e}^{hx^{32}}(-2+h^2 (x^{32})^2)$ and ${\rm e}^{h x^{32}}(x^{32}-h(x^{32})^2)$ constitute a basis of solutions for \eqref{fundODEn3} (this is true for $h\neq 0$ while for $h=0$ a basis is given by $1, x^{32}, (x^{32})^3, (x^{32})^4$).
Notice that the two solutions $1$ and $x^{32}$ are obvious since for $\e=-1$, $F$ and its first derivative do not appear in \eqref{fundODEn3}. Again, to determine the functions $f_1, \dots, f_4$ appearing in the expression above for $F$ it is necessary to substitute back that expression into system \eqref{fund2sysn3} and perform several computations. We obtain for $h\neq 0$:
\beq
\begin{split}
F(x^{21}, x^{32})=c_0\left(\frac{1}{3}hx^{32}+1-\frac{1}{6}h^2 x^{21}x^{32}-\frac{1}{3}hx^{21}\right)\\+c_1{\rm e}^{-hx^{21}}\left(6+4hx^{21}+h^2 (x^{21})^2+2hx^{32}+h^2x^{32}x^{21} \right)\\
+c_2\f{{\rm e}^{hx^{32}}\left( 6+h^2 (x^{32})^2+h^2x^{32}x^{21}-2hx^{21}-4hx^{32}\right)}{h},
\end{split}
\eeq
where $c_0, c_1, c_2$ are arbitrary constants. Notice that this solution is valid for $h\neq0$, since for $h=0$ the coefficient of $c_2$ is not defined.  This is consistent with the previous observation on the basis of solutions for \eqref{fundODEn3}. The corresponding $A$ can be recovered from the above formula for $F$ using \eqref{recoverA}. 

In the particular case $\epsilon=-1$, $h=0$, it is possible to find directly the general solution of system \eqref{fund2sysn3}. Indeed, in this case system \eqref{fund2sysn3} reduces to
\beq\label{fund2sysn32}
\begin{split}
x^{21}\f{\p^2 F}{\p x^{32}\p x^{21}}-x^{21}\f{\p^2 F}{(\p x^{21})^2}+2\f{\p F}{\p x^{21}}-\f{\p F}{\p x^{32}}=0,\\
-x^{32}\f{\p^2 F}{(\p x^{32})^2}+x^{32}\f{\p^2 F}{\p x^{32} \p x^{21}}+2 \f{\p F}{\p x^{32}}-\f{\p F}{\p x^{21}}=0,\\
x^{32}\f{\p^2 F}{\p x^{32} \p x^{21}}+x^{21}\f{\p^2 F}{\p x^{32} \p x^{21}}- \f{\p F}{\p x^{32}}- \f{\p F}{\p x^{21}}=0.
\end{split}
\eeq
A three parameter family of solutions determined using ${\tt Maple^{TM}}$ is given by 
$$F(x^{21}, x^{32})=c_1+c_2\left(x^{32}(x^{21})^3+\f{1}{2}(x^{21})^4\right)+c_3\left(\f{1}{12}(x^{32})^4+\f{1}{6}x^{21}(x^{32})^3 \right),$$
where $c_1, c_2, c_3$ are arbitrary constants. The corresponding $A$ can be recovered using \eqref{recoverA}.

\section{Examples preserving the bi-flat $F$-manifold structure}\label{examplesbiflat}
We have seen in Theorem \ref{theorembiflat} that reciprocal transformations preserving the flatness of both connections are homogeneous flat coordinates satisfying the further condition $e(A)=0$. Due to a Theorem proved in \cite{LP},
 in the case of the $\epsilon$-system with $n$ components there exist $n-1$ independent flat coordinates of this kind.

Let us consider the case $n=3$, $\epsilon\neq \f{1}{3}$. The computations will determine two of the three flat coordinates (the missing one is always given in any dimension by $\sum_{k=1}^n u^k$; it is a homogeneous function but it does not satisfy the condition $e(A)=0$). Although flat coordinates were computed also in \cite{LP}, the method followed here is computationally more straightforward, leading directly to hypergeometric equation. 

The general solution of the equations
$$e(A)=0, \quad E(A)=(1-3\epsilon)A,$$
can be written as 
$$A(u^1, u^2, u^3)=F\left(\f{u^3-u^2}{u^2-u^1}\right)(u^2-u^1)^{1-3\epsilon},$$
where $F$ is an arbitrary smooth function of one variable. (In general, the right degree of homogeneity of $A$ is given by $(1-\epsilon n)$ for the $\epsilon$-system in dimension $n$, with $\epsilon\neq \f{1}{n}$, see \cite{LP}.) It is convenient to define as before $x^{32}:=u^2-u^2$ and $x^{21}:=u^2-u^1$ and $z:=\f{x^{32}}{x^{21}}$. 
Now if this expression of $A$ is substituted into the three PDEs that express the fact that $A$ is a density of conservation laws, all three PDEs reduce to the {\em same} ODE of the second order in the unknown function $F$:
\beq\label{epsilonn3}
(1+z)\f{d^2F(z)}{dz^2}+\left(4\e+\f{2\e}{z} \right)\f{dF(z)}{dz}+\left(\f{3\e^2}{z}-\f{\e}{z}\right)F(z)=0,
\eeq
whose general solution can be expressed in terms of hypergeometric functions as follows:
$$F(z)=c_1{\rm hypergeom}([\e, 3\e-1], [2\e], 1+z)+c_2(1+z)^{1-2\e}{\rm hypergeom}([\e, -\e+1], [2-2\e], 1+z),$$
where $c_1$ and $c_2$ are arbitrary constants. 
Choosing $c_1=1, c_2=0$ and $c_1=0, c_2=1$ alternatively one obtains  two densities of conservation laws $A^1$ and $A^2$ that are also flat coordinates:
$$A^1(u^1, u^2, u^3)=(u^2-u^1)^{1-3\epsilon}\,{\rm hypergeom}\left([\e, 3\e-1], [2\e],\f{u^3-u^1}{u^2-u^1}\right),$$
$$A^2(u^1, u^2, u^3)=(u^2-u^1)^{1-3\epsilon}\,\left(\f{u^3-u^1}{u^2-u^1}\right)^{1-2\e}\,{\rm hypergeom}\left([\e, -\e+1], [2-2\e], \f{u^3-u^1}{u^2-u^1} \right).$$
In general it appears to be computationally very difficult to determine the corresponding currents for general values of $\epsilon$. 
As an example, we provide the current corresponding to $A^1$ in the case $\epsilon=1$. In this case we have
$$B^1(u^1, u^2, u^3)=\f{u^1+u^3}{(u^1-u^3)(u^2-u^1)}+\f{1}{u^1-u^3}+c_3,$$
where $c_3$ is an arbitrary constant. 

{\bf Acknowledgements} A. Arsie thanks the University of Toledo for financial support through selected funds while this work was completed and the Department of Mathematics and Applications of the University of Milan - Bicocca for the kind hospitality and the stimulating environment. 


\end{document}